\documentclass{sig}
\usepackage{epsfig}
\usepackage{epstopdf}
\usepackage{colortbl}
\usepackage{subfigure}
\usepackage{booktabs}
\usepackage{multirow}
\usepackage{slashbox}
\usepackage[ruled,vlined]{algorithm2e}
\usepackage{algorithm2e}
\newcommand{\WHD}[1]{\textcolor{black}{{#1}}}
\newcommand{\PZ}[1]{\textcolor{black}{{#1}}}
\newcommand{\mypara}[1]{{\smallskip \noindent \bf #1}\hspace{0.1in}}

\begin{document}
\conferenceinfo{IMC'15,}{October 28--30, 2015, Tokyo, Japan.}
\CopyrightYear{2015} 
\crdata{978-1-4503-3848-6/15/10}  

\title{Understanding Mobile Traffic Patterns of Large Scale Cellular Towers in Urban Environment}

\author{
      \alignauthor Huandong Wang$^{1}$, Fengli Xu$^{1}$, Yong Li$^{1}$, Pengyu Zhang$^{2}$, Depeng Jin$^{1}$\\
      \affaddr{$^{1}$Tsinghua National Laboratory for Information Science and Technology\\
Department of Electronic Engineering, Tsinghua University, Beijing 100084, China\\ $^{2}$University of Massachusetts Amherst}\\
     \email{\{liyong07, jindp\}@tsinghua.edu.cn, pyzhang@cs.umass.edu}
}

\maketitle
\begin{abstract}

Understanding mobile traffic patterns of large scale cellular towers in urban environment is extremely valuable for Internet service providers, mobile users, and government managers of modern metropolis.
This paper aims at extracting and modeling the traffic patterns of large scale towers deployed in a metropolitan city.
To achieve this goal, we need to address several challenges, including lack of appropriate tools for processing large scale traffic measurement data, unknown traffic patterns, as well as handling complicated factors of urban ecology and human behaviors that affect traffic patterns. Our core contribution is a powerful model which combines three dimensional information (time, locations of towers, and traffic frequency spectrum) to extract and model the traffic patterns of thousands of cellular towers. Our empirical analysis reveals the following important observations. First, only five basic time-domain traffic patterns exist among the 9,600 cellular towers. Second, each of the extracted traffic pattern maps to one type of geographical locations related to urban ecology, including residential area, business district, transport, entertainment, and comprehensive area. Third, our frequency-domain traffic spectrum analysis suggests that the traffic of any tower among the 9,600 can be constructed using a linear combination of four primary components corresponding to human activity behaviors. We believe that the proposed traffic patterns extraction and modeling methodology, combined with the empirical analysis on the mobile traffic, pave the way toward a deep understanding of the traffic patterns of large scale cellular towers in modern metropolis.

\end{abstract}




\section{Introduction}

The past few years have seen a dramatic growth in cellular network traffic, contributed by billions of mobile devices as the first-class citizens of the Internet. The global cellular network traffic from mobile devices is expected to surpass 24 exabytes ($10^{18}$) per month by 2019 \cite{cisco_cellular_forcast}, 9$\times$ larger than the traffic served by existing cellular network. While we are embracing a world with ambient cellular connectivity, however, we are facing a critical and challenging problem --- we have limited understanding about the patterns of traffic experienced by cellular towers deployed in urban areas, especially when 3G and LTE networks are widely available in current modern metropolis\cite{cisco_cellular_forcast,cici2015decomposition,das2014contextual}.
We do not completely understand how urban functional regions and ecologies, such as business district, affect the mobile traffic of cellular towers\cite{cici2015decomposition}. In addition, the dominant factors that affect their traffic variations
are still unknown. Such limited knowledge significantly increases the cost of operating thousands of cellular towers in big cities.

Despite of the aforementioned lack of knowledge, understanding the traffic patterns of cellular towers in the large scale urban environment is extremely valuable for Internet service providers (ISP), mobile users, and government managers of modern cites\cite{dong2014inferring,lee2014spatial,shafiq2012characterizing}. If we can {\em identify and model} the patterns of cellular towers, instead of using the same strategy to provide services, such as using the same load balancing and data pricing algorithms on each tower, an ISP can exploit the modeled traffic patterns and customize the strategies for individual cellular towers. For example, an ISP can potentially have different pricing on individual cellular tower based on the traffic it experiences. In addition, mobile users will benefit from the traffic modeling as well because they can choose towers with predicted lower traffic and enjoy better services. Surprisingly, management departments of government will benefit from the traffic modeling as well because they may infer the land usage and human economy activities by looking at the patterns of cellular traffic\cite{kosinski2013private}.

On the other hand, understanding the traffic patterns of cellular towers is challenging for three reasons. First, the traffic experienced by thousands of cellular towers deployed in large scale modern cities is complicated and hard to analyze. For example, our dataset includes 9,600 cellular towers and 150,000 subscribers, where lots of redundant and conflict logs are observed. To identify traffic patterns embedded in the thousands of towers, we need to design a system that is able to clean and handle the data of large scale cellular traffic. Second, we do not have the priori about the existence of patterns that can be used for representing the behavior of thousands of cellular towers. To make matters worse, even if such patterns exist, we do not know their profiles. Without these profiles, it is challenging to group thousands of cellular towers into a small number of patterns. Third, the traffic of a cellular tower is affected by many factors, such as time and locations, etc. These factors, sometimes, compound with each other and further complicate our analysis. For example, significant traffic variation is observed at both fine-grained (hours) and coarse-grained (days) time scale, and across towers deployed in different locations\cite{yongli2015hotplant,lee2014spatial}. By addressing these challenges, in this paper, we investigate how to extract and model the mobile traffic patterns of thousands of cellular towers in a large scale urban environment via credible dataset collected by one of the largest commercial mobile operators.

Our core contribution is a powerful model which combines three dimensional information, including time, locations of towers, and traffic frequency spectrum, for extracting and modeling the traffic patterns of thousands of cellular towers. A breakdown of the core contribution comprises three parts. First, we design a system which leverages machine learning to identify and extract five patterns from the traffic of thousands of cellular towers. Our system is built with processing large scale data in mind and is able to process the traffic of \WHD{thousands of} towers with  granularity of 10 minutes. Second, we identify the \WHD{geographical} context of traffic experienced by cellular towers by investigating the correlation between time-domain traffic characteristics and \WHD{geographical} locations of towers. Therefore, by looking at the traffic pattern of a tower, we can infer the type of location where it is deployed and the type of users it serves. Third, our frequency-domain traffic spectrum analysis reveals that any traffic of the 9,600 cellular towers can be constructed using a linear combination of four primary components corresponding to human activity behaviors. This observation provides an unique angle (frequency) for analyzing cellular traffic and significantly simplifies the process of analysis by a linear model.

Through investigating the traffic of 9,600 cellular towers, we find following interesting observations. First, the 9,600 cellular towers can be classified into five groups using features extracted from time-domain traffic. This experimental result confirms our motivation that a small number of patterns do exist among thousands of cellular towers. Second, each of the traffic pattern maps to one type of \WHD{geographical} locations, including resident, office, transport, entertainment, and comprehensive area. Therefore, the traffic pattern of a cellular tower does suggest the urban ecology and \WHD{geographical} location context where it is deployed as well as the type of users it serves. Third, our frequency-domain analysis reveals that the transition between the five traffic patterns encodes the mobility of human. For example, when the phase of residential pattern moves toward the phase of transport pattern, people start their commute from home to work. In summary, we believe that the proposed traffic patterns extraction and modeling, combined with the empirical study on large scale cellular towers, pave the way toward a deep understanding of the traffic patterns of large scale cellular towers.

This paper is structured as follows. In Section 2, we provide details about the utilized dataset, and present some basic observations of traffic spatio-temporal distributions. In Section 3, we design our traffic processing system and identify the key traffic patterns of the large scale cellular towers. Based on the discovered five traffic patterns, in Section 4 and 5, we conduct a deep analysis and reveal the correlation among data traffic, urban ecology and human behaviors in the time and frequency domain respectively.  After discussing related work in Section 6, we summarize our discoveries and discuss potential investigations in Section 7.

\section{Dataset and visualization}
In this section, we provide details about the dataset we investigate as well as the needed preprocessing. In addition, we visualize the spatial-temporal distribution of cellular traffic.

\subsection{Dataset Description}
The dataset is an anonymized cellular trace collected by an ISP from Shanghai, a big city in China, between Aug 1st and Aug 31st 2014. Each entry of the trace contains detailed mobile data usage of \WHD{150,000} users, including the ID of devices (anonymized), start-end time of data connection, base station ID, address of base station, and the amount of 3G or LTE data used in each connection. The trace logs 1.96 billion tuples of the described information, \WHD{contributed by approximately 9,600 base stations all over Shanghai}. The trace contains \WHD{2.4} petabytes ($10^{15}$) logs, \WHD{77} terabytes ($10^{12}$) per day and \WHD{8} gigabytes ($10^{9}$) per base station on average. This large scale and fine-grained dataset guarantees the credibility of our traffic pattern analysis and modeling.

\begin{figure*}[t]
\centering
\subfigure[Hourly]{\includegraphics[width=.31\textwidth]{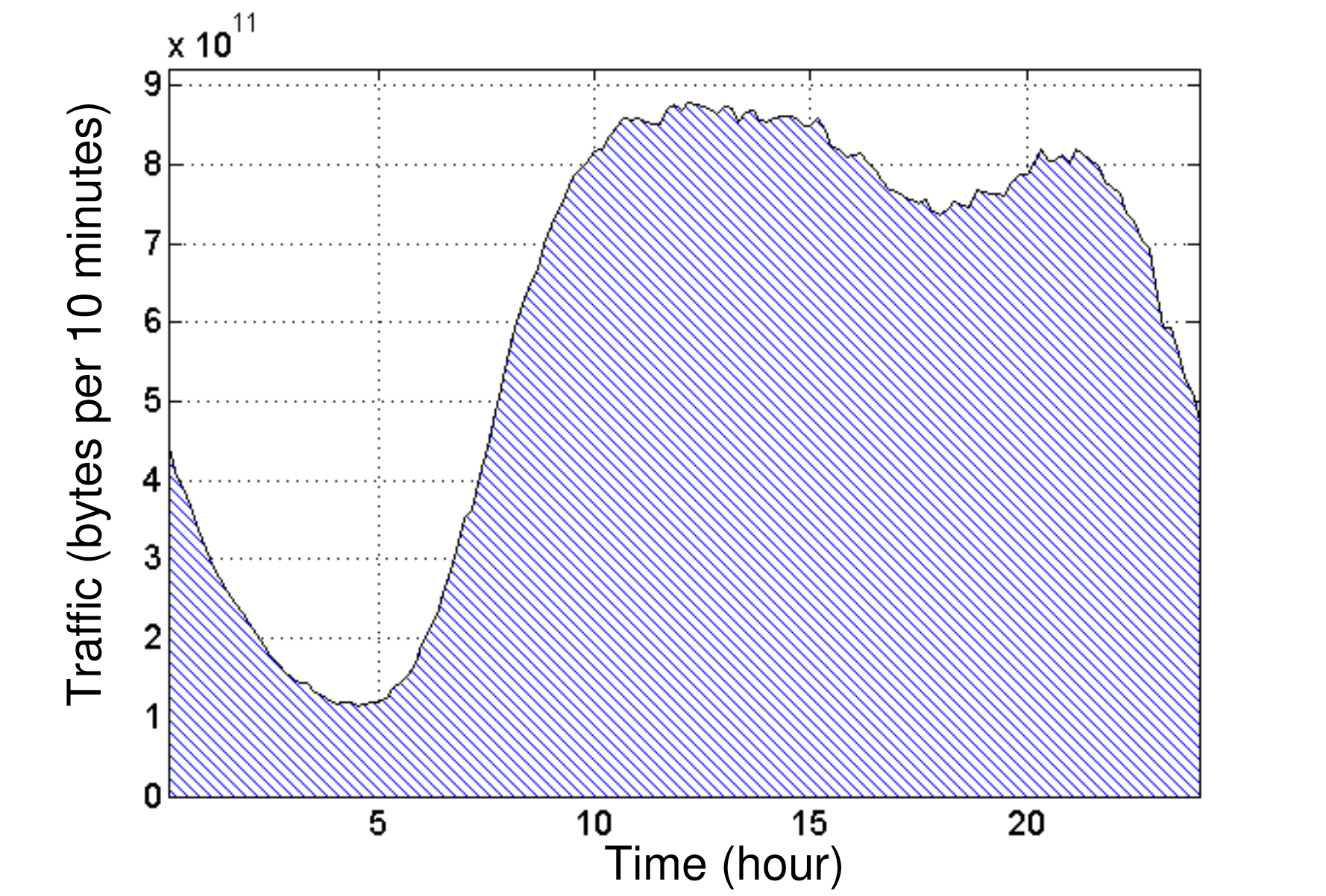}}
\subfigure[Daily]{\includegraphics[width=.31\textwidth]{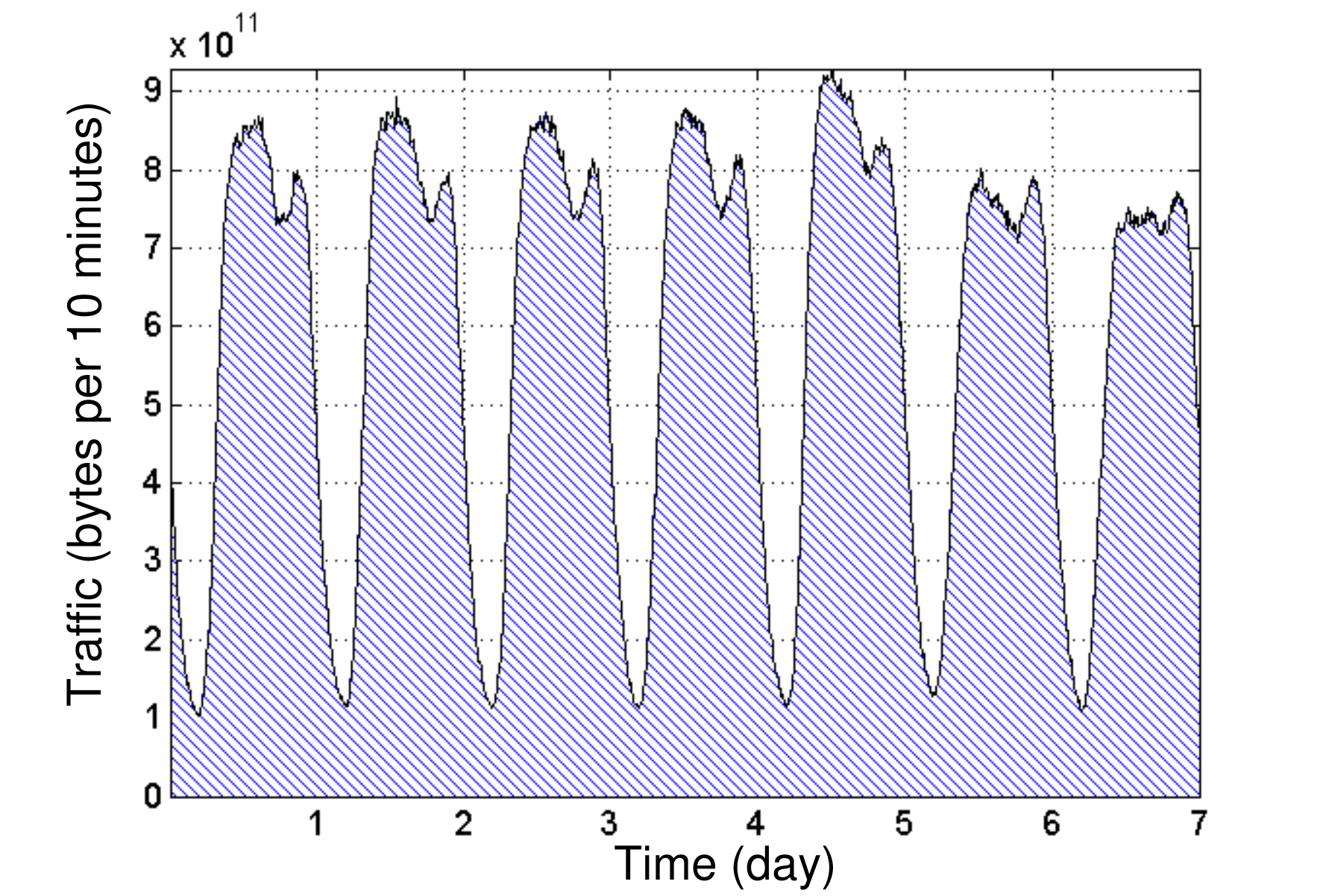}}
\subfigure[Weekly]{\includegraphics[width=.31\textwidth]{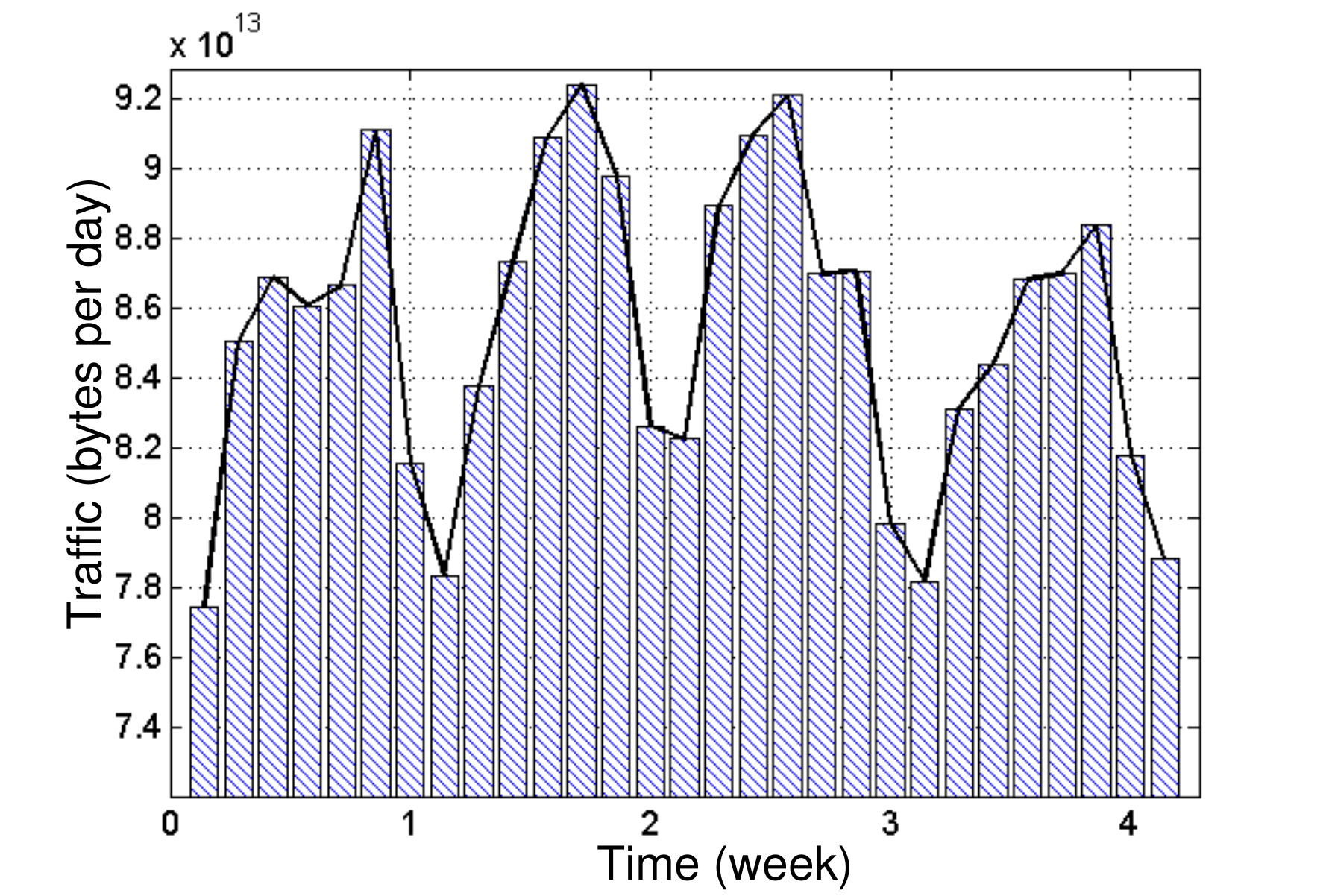}}
\caption{\WHD{The temporal distribution of cellular traffic at different time scales.}} \label{TD}
\end{figure*}

\subsection{Preprocessing}
The trace collected by the ISP needs to be preprocessed because of \PZ{the existence of redundant and conflict traffic logs as well as the incomplete information of base stations' locations}. The preprocessing includes three steps. First, we eliminate the redundant and conflict logs, such as the identical traffic logs, introduced by technical issues.
Second, to solve the problem of incomplete information, we convert the addresses of  base stations to their \WHD{geographical} longitudes and latitudes through APIs provided by Baidu Map, the most popular online map service provider in China.
This conversion gives us the precise \WHD{geographical} location of a base station, which is important for analyzing the ground truth of urban functional regions. The last step of preprocessing is computing the traffic density (byte/km$^2$) across the city. The obtained traffic density allows us to understand the spatial distribution of cellular traffic.

\subsection{Data Visualization}

Before diving into a deep analysis of mobile data traffic, we first visualize the spatial-temporal traffic distribution of the 9,600 base stations, where we find two interesting observations.

First, the data embeds fundamental temporal patterns of mobile data traffic. Figure~\ref{TD} shows the aggregated traffic of the 9,600 towers at different time scales. Figure~\ref{TD}(a) shows the traffic distribution of a day (Aug 7th \WHD{2014}, Thursday) where we observe that the aggregated network traffic is tightly coupled with the sleep pattern of humans. High cellular traffic is observed during the day and low traffic is experienced during midnight. There are two traffic peaks in each day: one around 12PM and the other around 10PM. Similar patterns are observed in Figure~\ref{TD}(b). The timing of the two peaks suggests that most people tend to consume data traffic heavily after lunch and before sleep. Figure~\ref{TD}(b) shows the traffic distribution of a week
(from Aug \WHD{4} to Aug \WHD{10} \WHD{2014})
and Figure~\ref{TD}(c) shows the traffic distribution of a month (from Aug 3 to Aug 31 \WHD{2014}). Both figures show that the traffic exhibits a periodical pattern on the scale of a week, where weekend's traffic is less than weekday's traffic. Such traffic variation comes from people's weekly work schedule.

\begin{figure} [tb]
\centering
\subfigure[4AM]{\includegraphics[width=.21\textwidth]{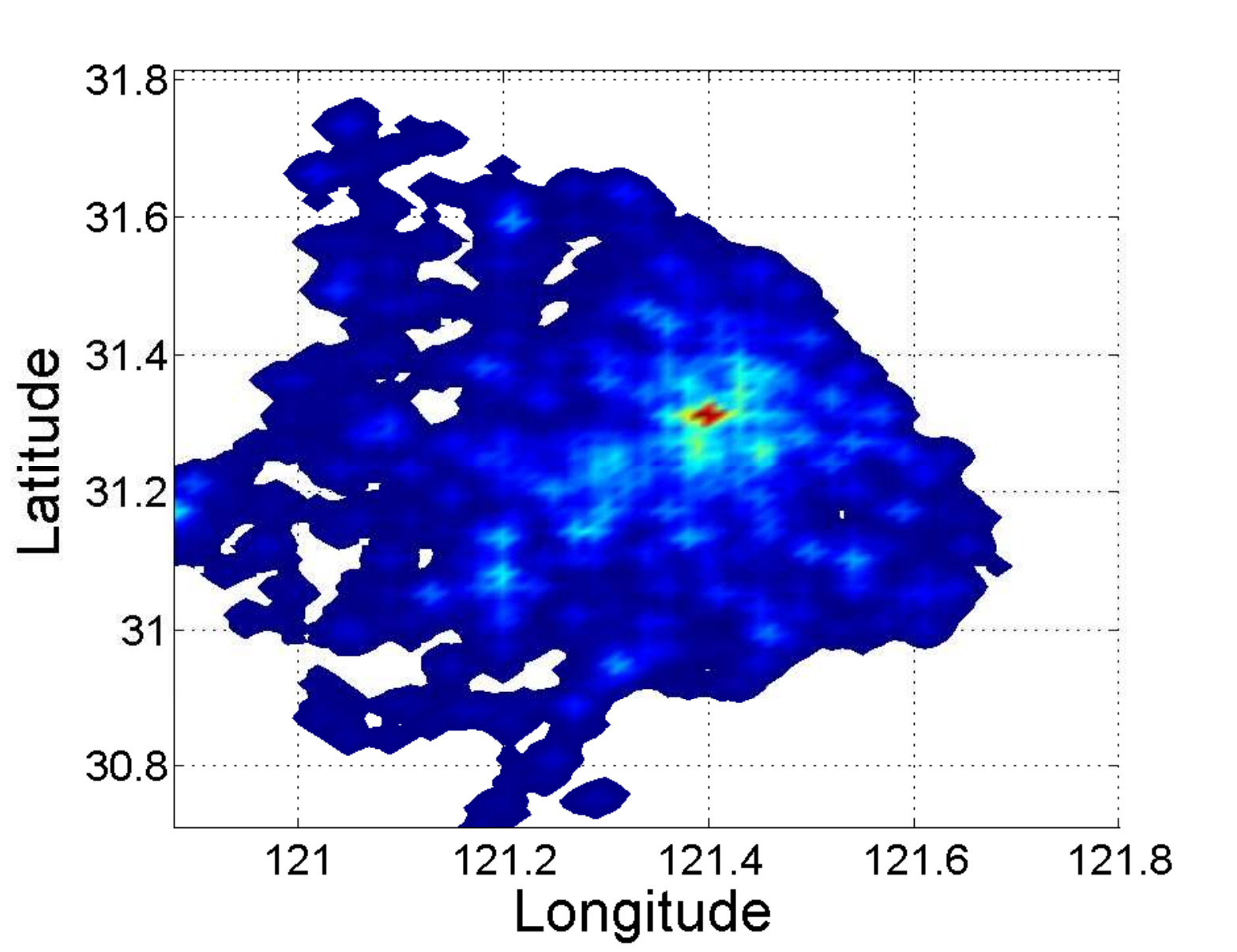}}
\subfigure[10AM]{\includegraphics[width=.237\textwidth]{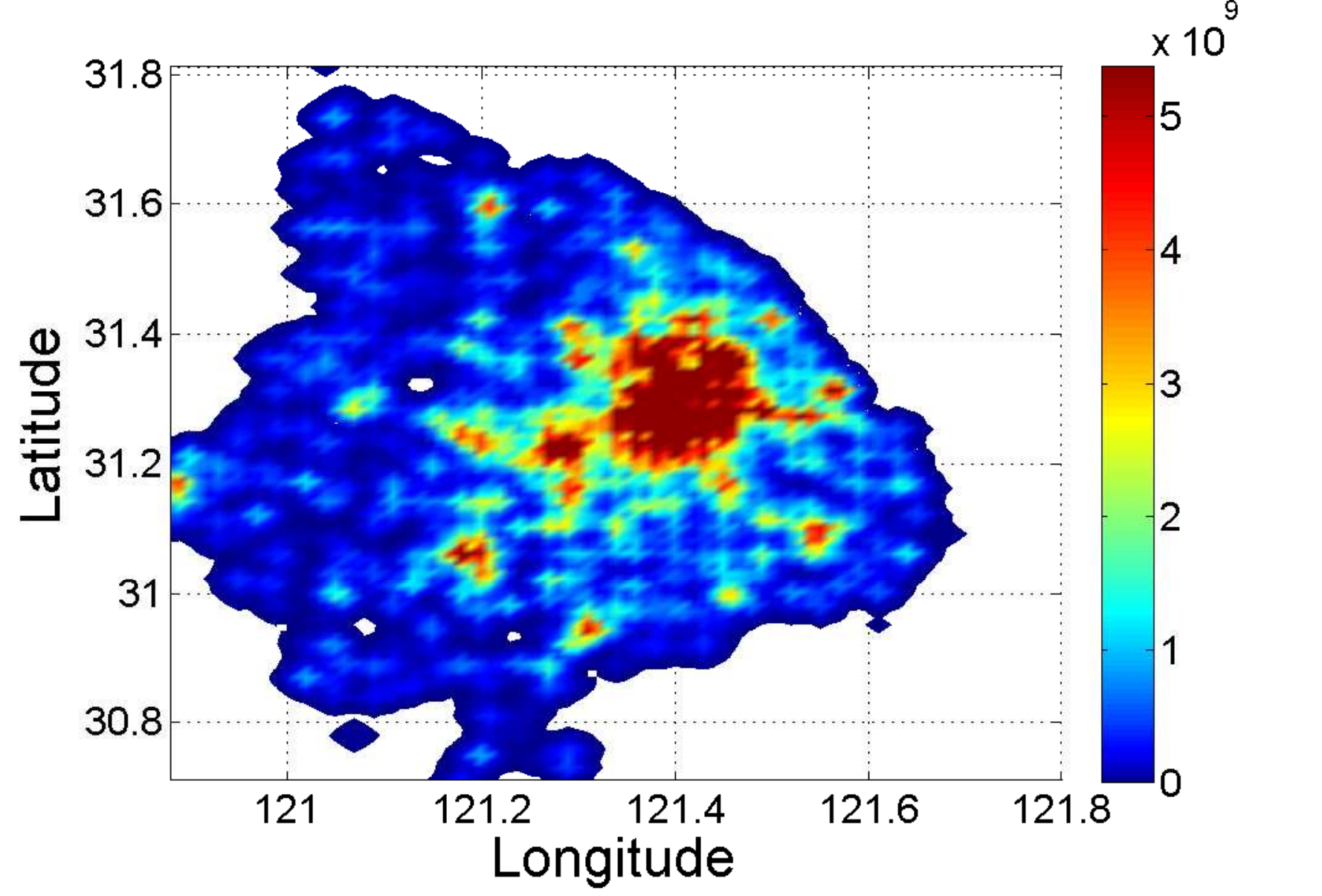}}
\subfigure[4PM]{\includegraphics[width=.21\textwidth]{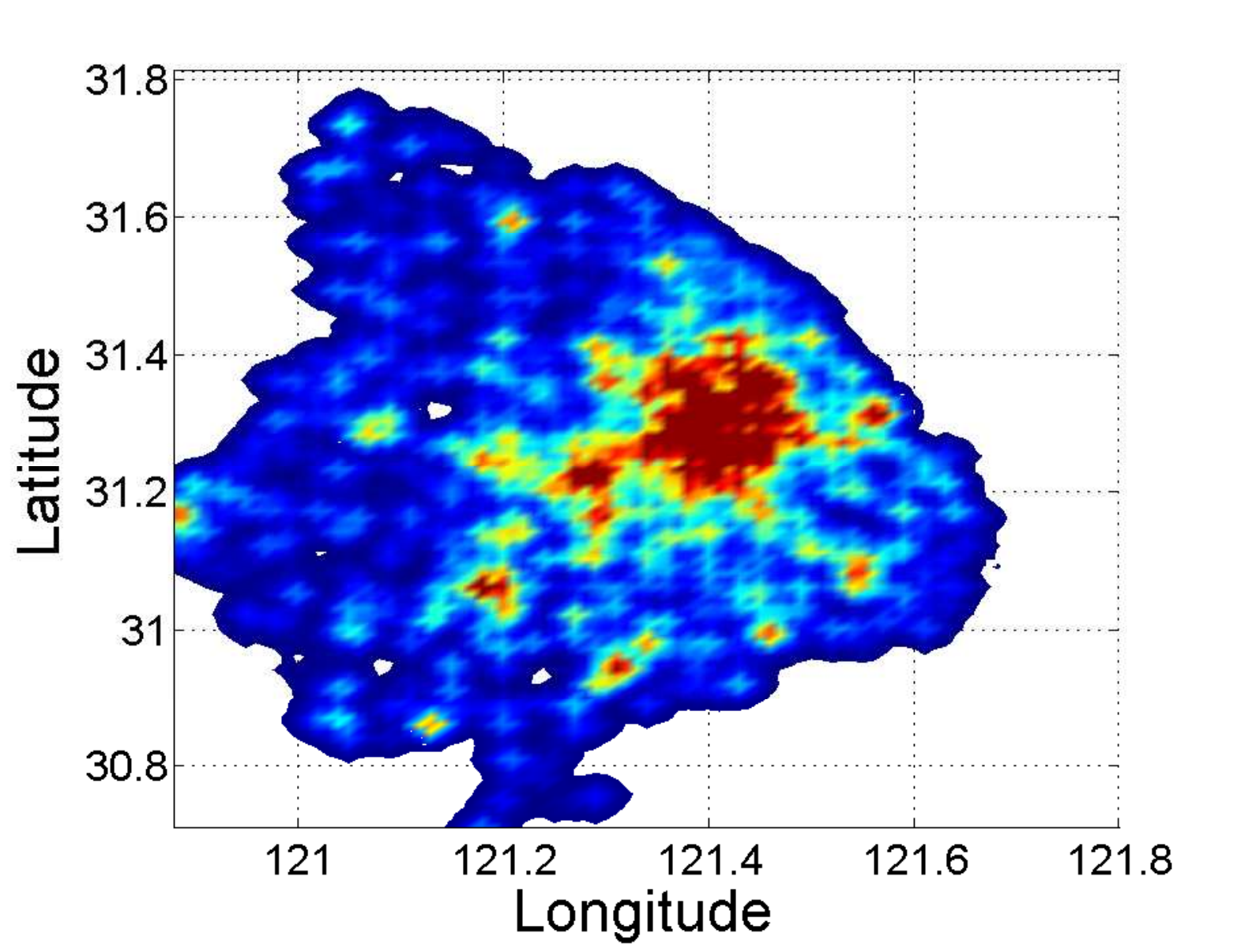}}
\subfigure[10PM]{\includegraphics[width=.237\textwidth]{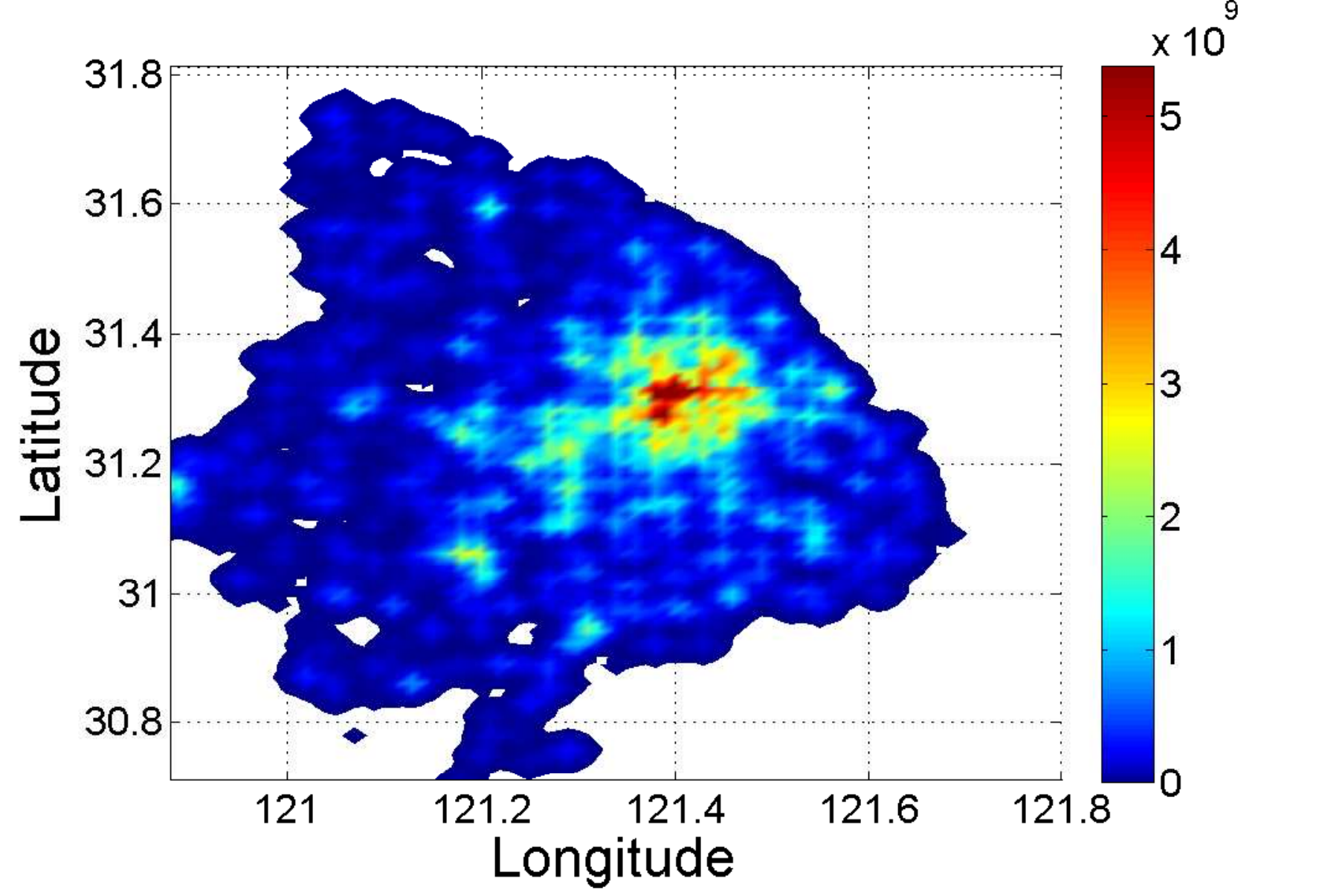}}
\caption{\WHD{The spatial distribution of cellular traffic at different time.}} \label{SDDT}
\end{figure}

On the other hand, our trace also records the spatial distribution of mobile data traffic. Surprisingly, we find that the spatial and temporal characteristics of traffic are correlated. Figure~\ref{SDDT} shows the \WHD{geographical} traffic density \WHD{(bytes transmitted per hour per km$^2$)} at 4AM, 10AM, 4PM and 10PM.
\WHD{As shown in the color bar, the red one indicates higher traffic and the blue one stands for lower traffic.}
We find the following observations. First, towers deployed at the center of the city experience high traffic despite of the time of a day. Second, at 4AM, most areas of the city are covered by dark color, which suggests that traffic demand is small because of human sleep. In contrast, at 10AM, most areas of the city are covered by light color, suggesting that traffic demand becomes high because people start working. Therefore, the areas of peak traffic map to areas occupied by human, such as residential housing or central business district (CBD).

\section{Identifying Traffic Patterns of Cellular Towers }
Now, we investigate the data traffic of the \PZ{thousands} of 3G/LTE cellular towers and design a system that is able to identify key traffic patterns of large scale cellular towers. We start from understanding the traffic patterns of a few cellular towers to motivate our study.

\subsection{Motivation and Problem Statement}
\begin{figure} [t]
\begin{center}
\includegraphics*[width=8.3cm]{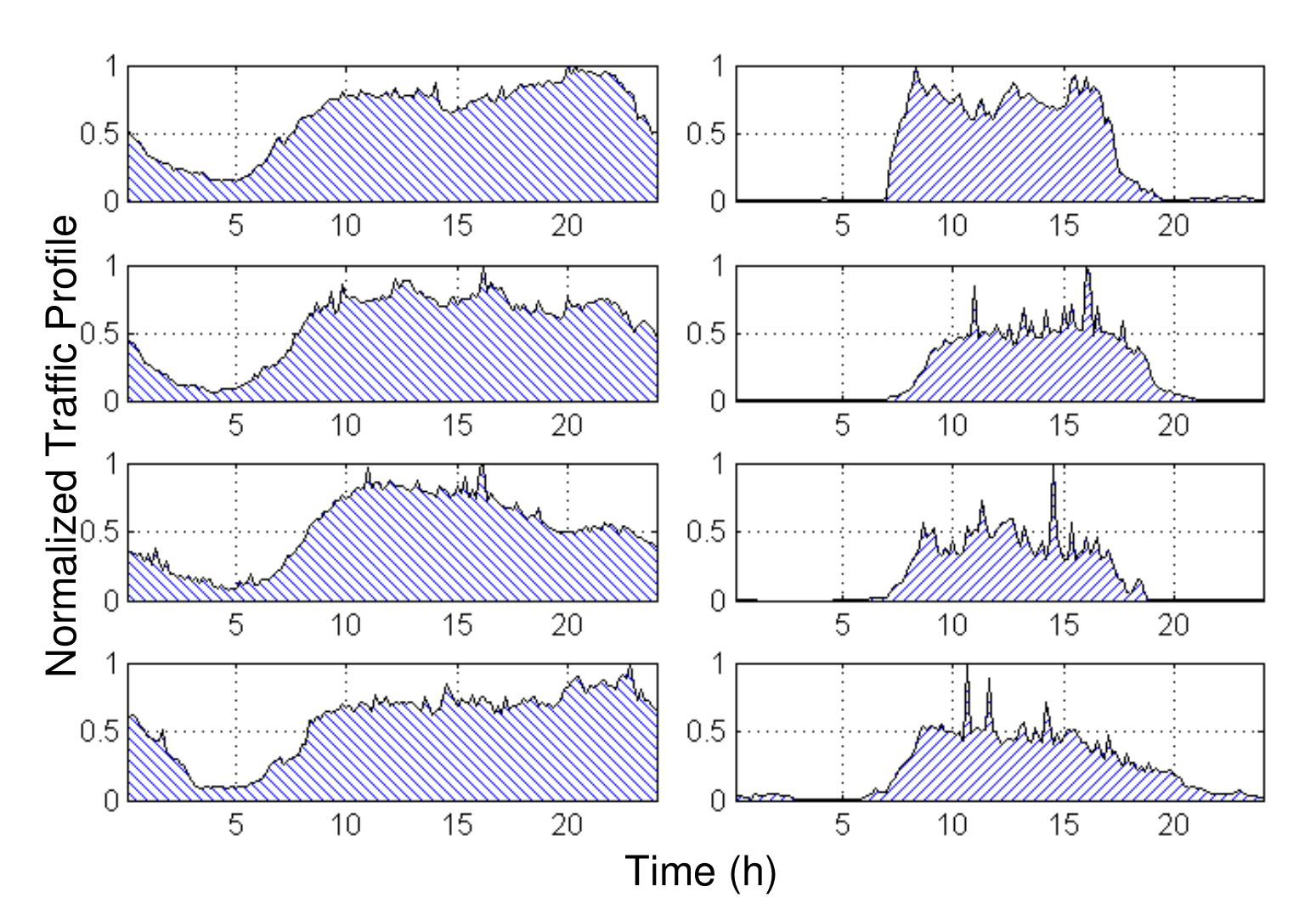}
\end{center}
\caption{Cellular traffic experienced by base stations deployed in the residential area and business district.} \label{fig:BSexam}

\end{figure}

Our cellular network traffic measurement and analysis are motivated by a key observation --- the traffic pattern of one cellular tower is vastly different from another. 
\WHD{Through online map service,
we randomly select four towers from the positions of residential areas and four towers from business districts, and plot their normalized traffic profile in the left and right column of Figure~\ref{fig:BSexam}, respectively.}
We can clearly observe the difference of traffic between these two types of cellular towers, where the traffic profiles of residential towers have two peaks within a day and remain high across night, while the traffic profiles of towers in business district experience only one peak within a day and get close to zero across night.
This comparison clearly reveals the difference of traffic patterns between the two specific types of cellular towers. However, from the perspective of an ISP, which manages \PZ{thousands} of cellular towers, is the traffic pattern of one cellular tower vastly different from another? To understand this problem, we conduct a large scale measurement and investigate the recorded 9600 cellular towers in our dataset of Shanghai.

\begin{figure}[t]
\centering
\subfigure[Latitudes]{\includegraphics[width=.220\textwidth]{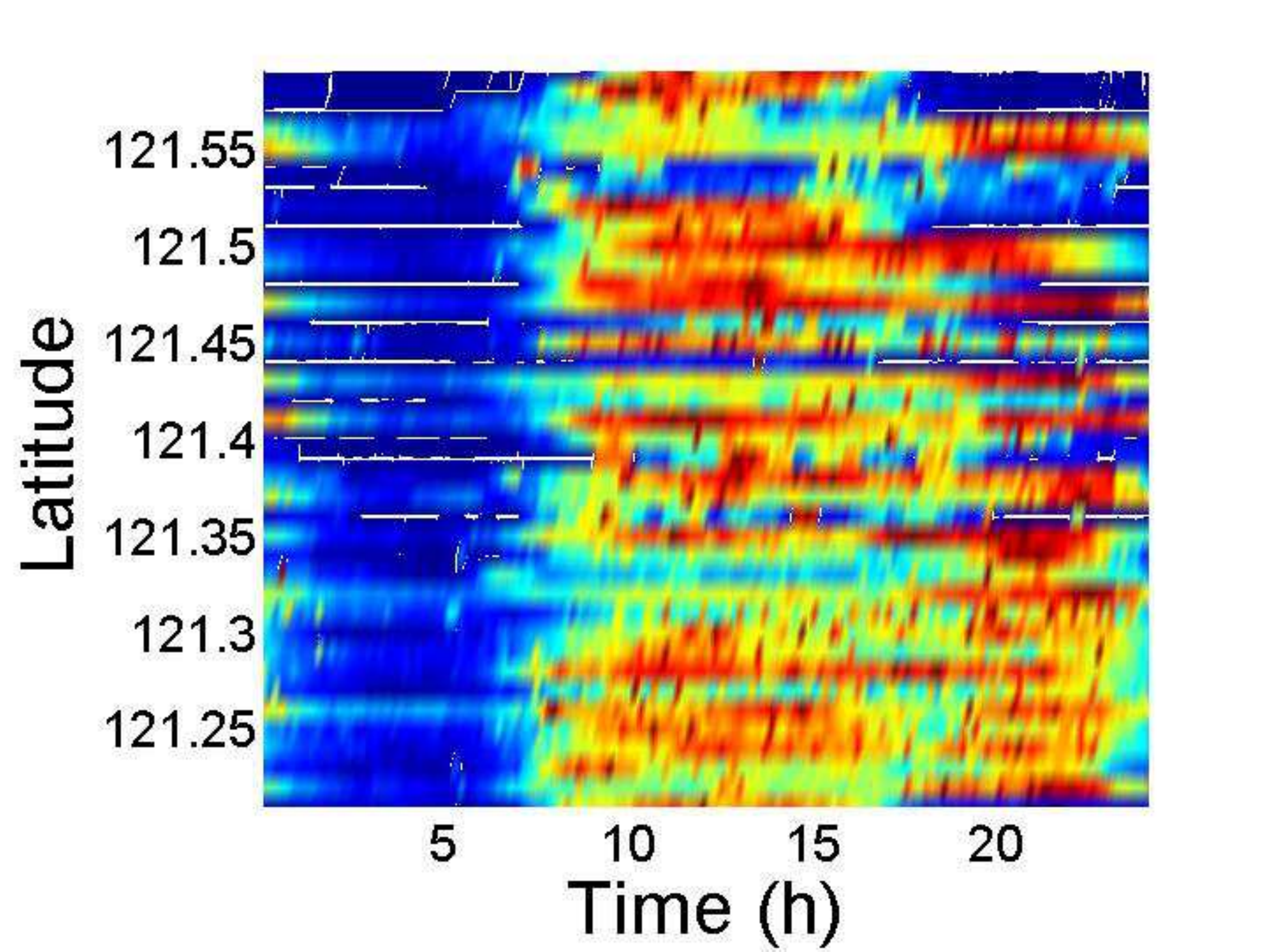}}
\subfigure[Longitudes]{\includegraphics[width=.250\textwidth]{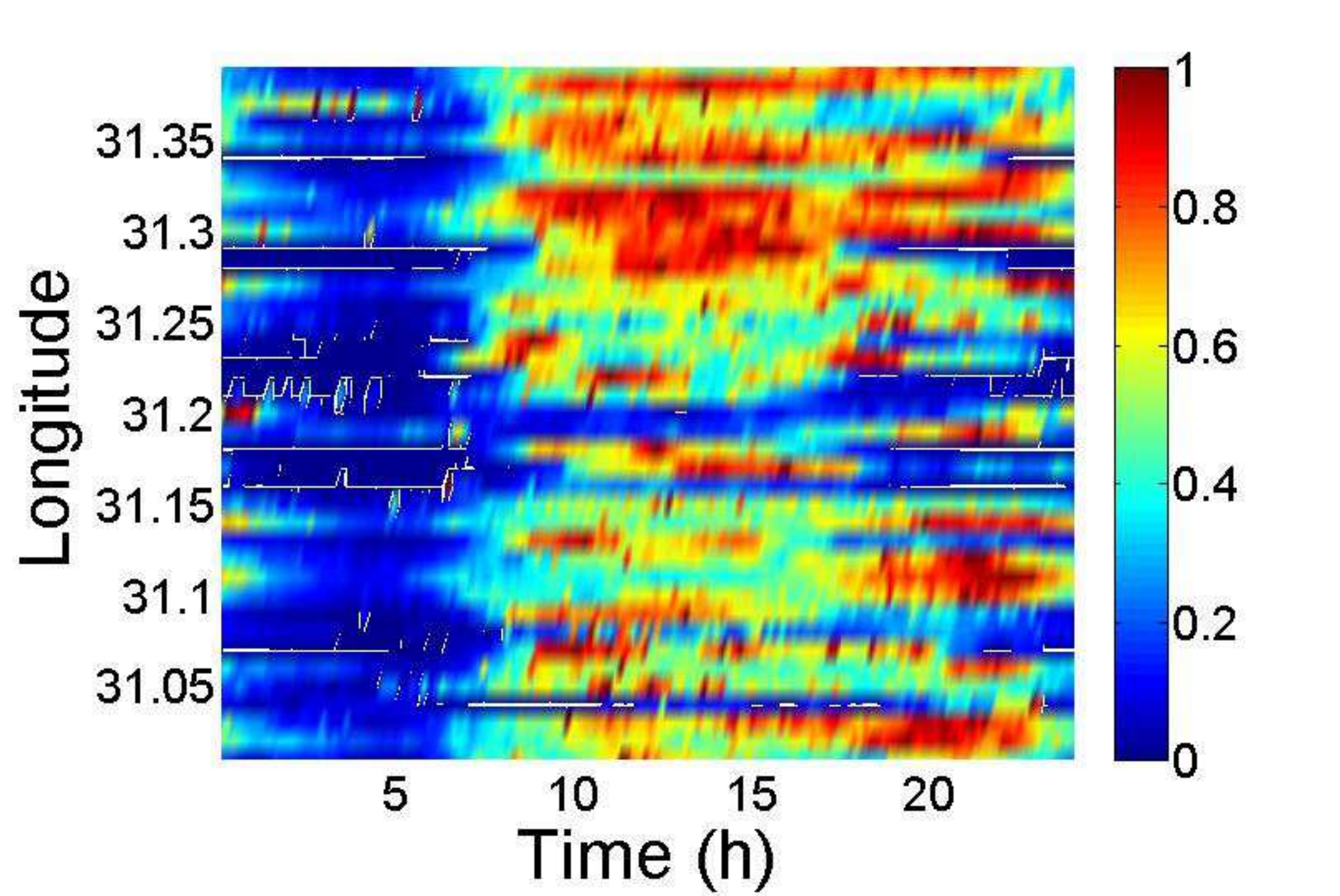}}
\caption{Cellular traffic experienced by base station randomly selected from different latitudes and longitudes. Large traffic variations are observed.} \label{fig:LATvsTime_all}

\end{figure}

Figure~\ref{fig:LATvsTime_all} shows the \WHD{normalized} traffic variations within one day with 40 randomly selected cellular towers for each \WHD{0.01} degree latitudes or longitudes respectively. The x-axis shows the time in hours and y-axis shows the logical positions of the selected cellular towers in terms of latitude (a) or  longitude (b).
\WHD{For example, the first row of pixels in Figure~\ref{fig:LATvsTime_all}(a) represent the traffic variations of one cellular towers, of which the latitude is around 121.60 and the longitude is randomly selected.}
\WHD{Traffic measured on each cellular tower is normalized by its maximum, }
and the color presents the normalized value where red color indicates higher traffic and blue color stands for lower traffic \WHD{as shown in the color bar.}
In these measurements, we find two observations. First, the peak hour of one cellular tower, which is marked as red, is vastly different from another during the day time when serving mobile users. In fact, the variance of the peak among the selected towers is about 10 hours. Second, while most of towers experience low traffic in early morning, the first several towers
\WHD{in Figure~\ref{fig:LATvsTime_all}(a), of which the latitude is around 121.60}, also have low traffic during evening. Therefore, significant differences of data traffic are observed across cellular towers. Such differences cause troubles for an ISP to manage its cellular network. For example, because of the unique pattern of individual traffic, an ISP cannot obtain the optimal performance by using the same load balancing strategy, which is built on top of traffic patterns, on different towers. Therefore, a natural question to ask is that is it possible to model the traffic pattern of \PZ{thousands} of cellular towers? More specifically, can we utilize a few simple patterns to present the traffic of \PZ{thousands} of cellular towers? Identifying these patterns of cellular towers would give an ISP significant benefits on network management, including load balancing, pricing, etc.

\begin{figure}[t]
\centering
\subfigure[Residential Area base stations]{\includegraphics[width=.220\textwidth]{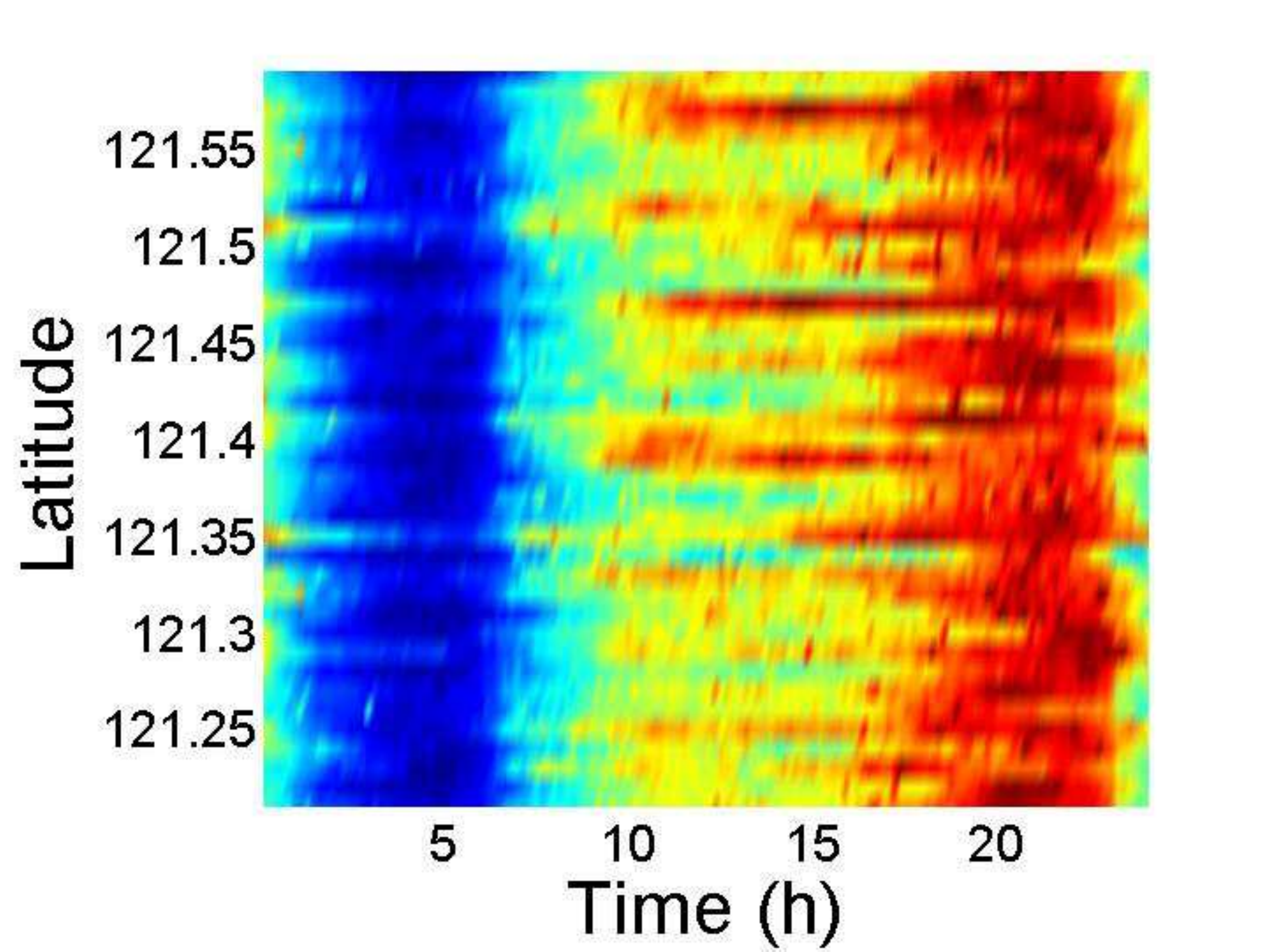}}
\subfigure[Business District base stations]{\includegraphics[width=.250\textwidth]{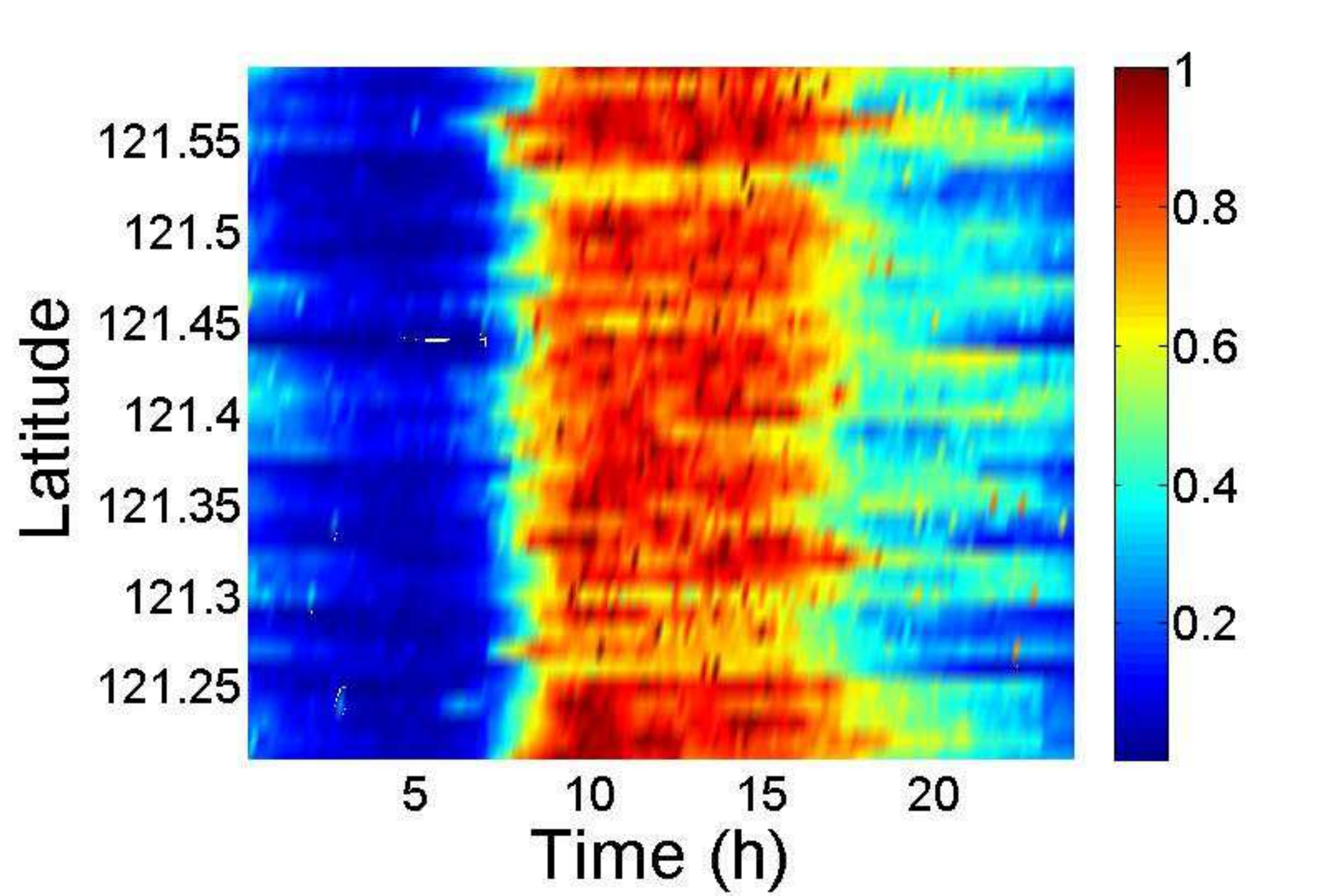}}
\caption{Cellular traffic experienced by base stations selected from residential and business district.} \label{fig:fig2patterns}

\end{figure}

Our investigation suggests that at least two, maybe more, traffic patterns exist among \PZ{thousands} of cellular towers. Figure~\ref{fig:fig2patterns} shows the \WHD{normalized} traffic \WHD{profile} of 40  
selected cellular towers deployed in residential area and in business district
\WHD{for each 0.01 degree latitudes}.
\WHD{Compared with the disorder in both temporal and spatial dimension exhibited in Figure~\ref{fig:LATvsTime_all},
traffic variations for cellular towers in a single kind of regions are more regular and similar to each other.}
In addition, we find other two observations in this investigation. First, in terms of the traffic of residential area, all residential towers experience similar traffic patterns where the peak traffic is present around 9PM. In addition, only a small amount of traffic is observed between 8AM and 4PM because most users leave home for work. Similar conclusion can be drawn for towers deployed in business district. Second, the traffic pattern of residential towers is different from towers deployed in business district where peak hour appears around 1PM. Inspired by these two observations, we conclude that traffic patterns do exist among \PZ{thousands} of cellular towers. One key question addressed by this paper is finding out how many traffic patterns exist among \PZ{thousands} of cellular towers and how to identify them.

\subsection{Identifying Traffic Patterns of Cell-towers}
Investigating traffic patterns among \PZ{thousands} of cellular towers is extremely challenging for three reasons. First, we have little prior knowledge about the data traffic, and do not know which cell towers may share the same traffic pattern and how the pattern may look like. Second, the measured cellular traffic data is huge in terms of tracing 9,600 cellular towers for a month. To make matters worse, the measured data is not clean in terms of unstructured logs. Last but not least, the measured cellular traffic data is noisy where large variation of traffic is observed because the absolute traffic depends on the number of mobile users served. All these factors make the analysis of cellular traffic patterns extremely challenging. To tackle these challenges, we design, implement, and evaluate a system which is able to identify the key traffic patterns of such large scale cellular towers. Our system is composed by three key elements: traffic vectorizer, pattern identifier and metric tuner.

\mypara{Traffic vectorizer:}
\PZ{We implement a traffic vectorizer on Hadoop platform to convert the large scale unstructured traffic logs into traffic usage vectors.}
The key of designing the traffic vectorizer is a parallel transformer, which takes the time-domain traffic logs of \PZ{thousands} of cellular towers as its input and converts each cell tower's logs into a time-domain traffic vector.
The vector is constructed in two phases --- \PZ{aggregation} and normalization. \PZ{In the first phase, each cellular tower's traffic logs are segmented into thousands of chunks, with each chunk contains 10-minutes traffic logs. Then we aggregate the traffic logs in each chunk and generate a traffic usage vector.}
\PZ{In the second phase, we perform zero-score normalization on each vector to eliminate their differences in amplitude, in order to find out the similar traffic patterns without the interference of different amplitude.}
\PZ{We define the traffic vector of cellular tower $j$ as $X_j\!=\!(x_j[1],...,x_j[N])^T$, with $x_j[i]$ stands for the normalized traffic amount in the $i_{th}$ 10-minute time slot.
We remove 3 days from the month to make the duration consist of four entire weeks.
Thus, $N$ is number of 28 days' 10-minutes segmentation, i.e., 4032 in our analysis.}

\mypara{Pattern identifier:} Pattern identifier takes the vectorized data from the vectorizer and runs an unsupervised machine learning algorithm for identifying the key patterns of cellular tower traffic. The pattern identifier addresses one key challenge of the mining process --- unknown patterns, by exploiting hierarchical clustering\cite{corpet1988multiple}. The basic idea of hierarchical clustering is iteratively merging the nearest two clusters. It first considers each input point as a cluster and then bottom-up iteratively merges the nearest two clusters until the stop condition is met. \PZ{In the clustering, we use the euclidean distance as the distance metric and define the distance between clusters as average-linkage distance. In addition, we set a threshold value as stop condition, which stops the clustering when the distance between two clusters is above the threshold value.}

\begin{figure}[t]
\centering
\subfigure[\PZ{DBI variation}]{\includegraphics[width=.235\textwidth]{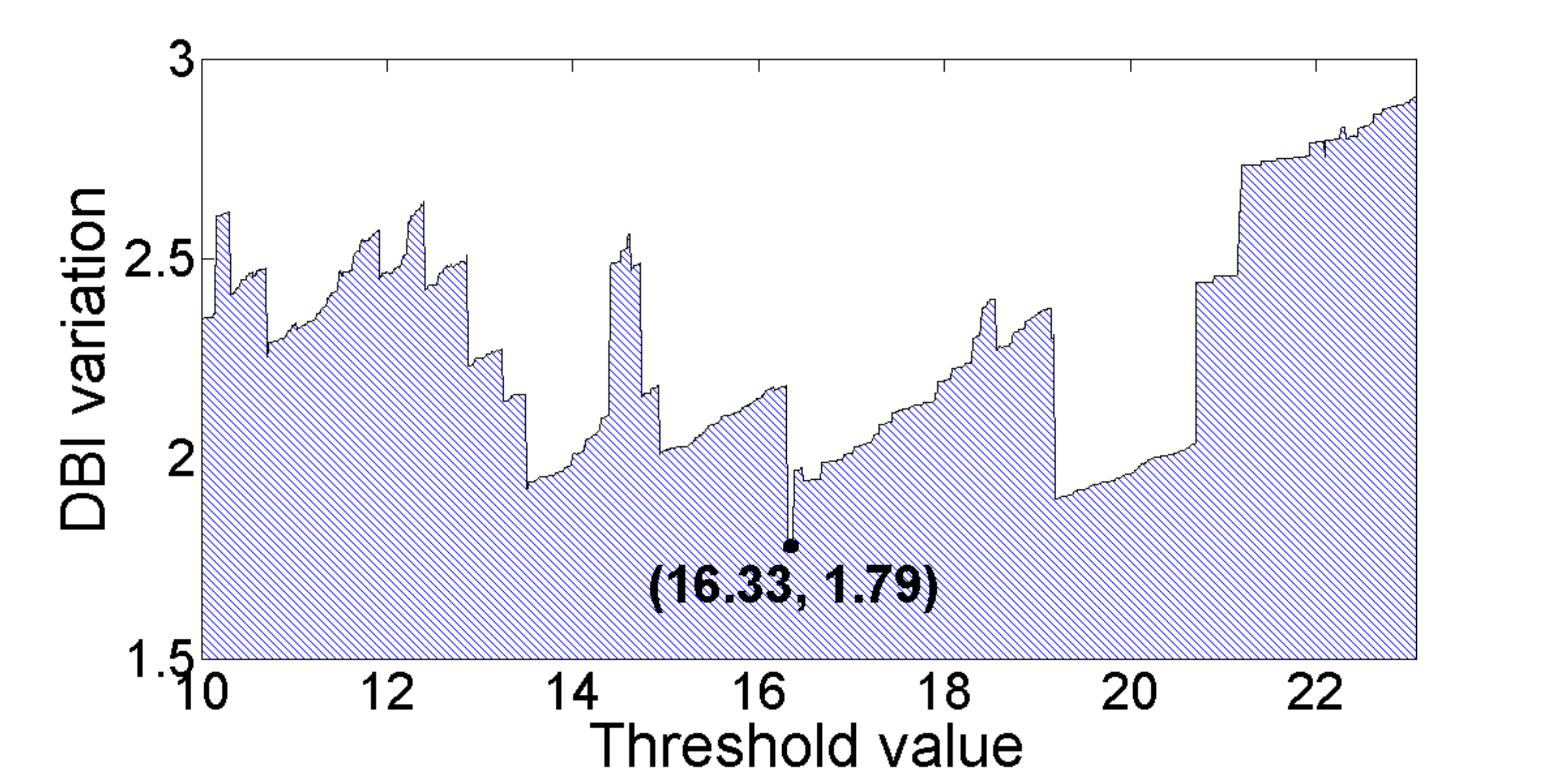}}
\subfigure[CDF of distance]{\includegraphics[width=.235\textwidth]{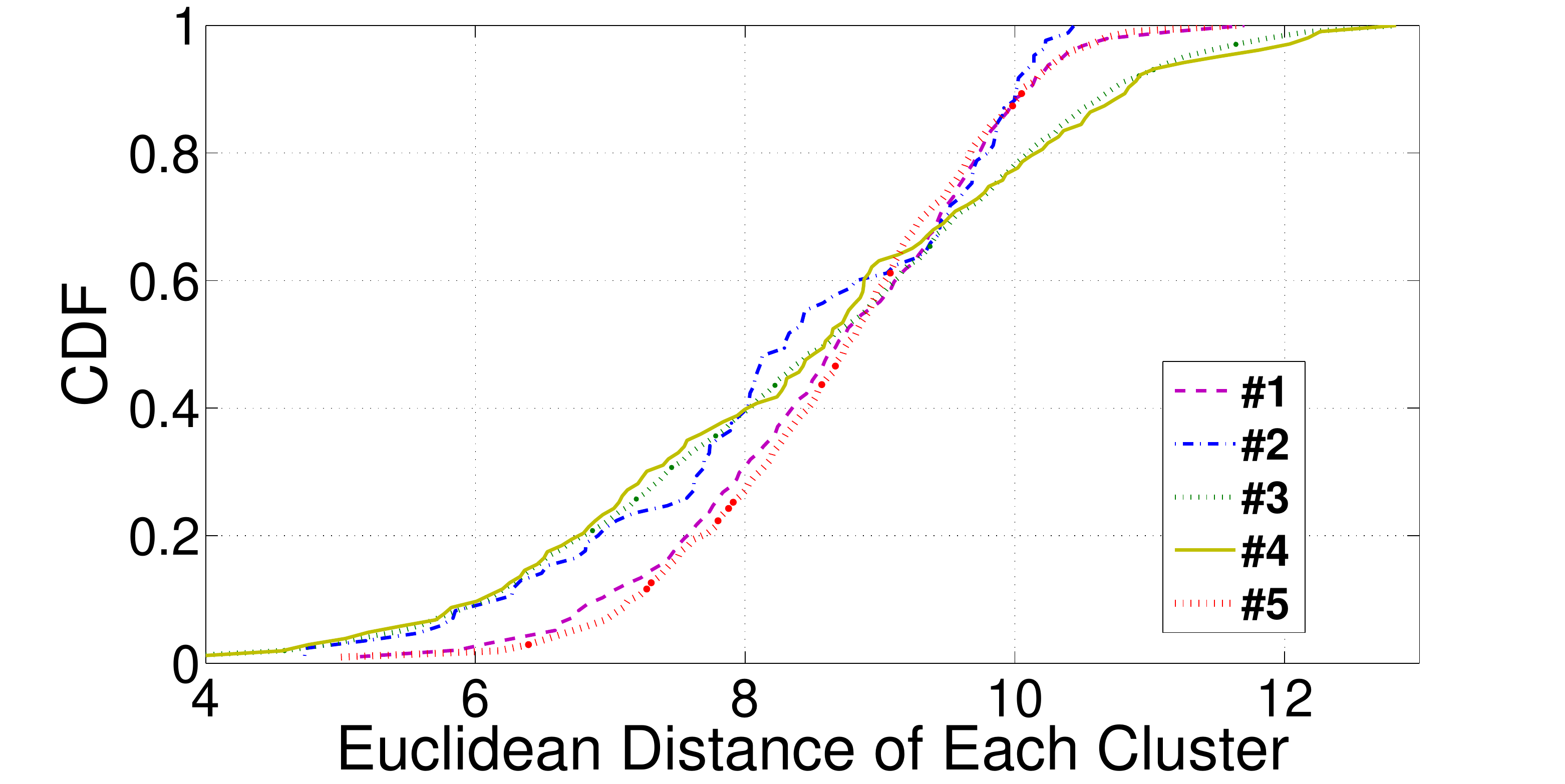}}
\subfigure[\#1:Resident area]{\includegraphics[width=.235\textwidth]{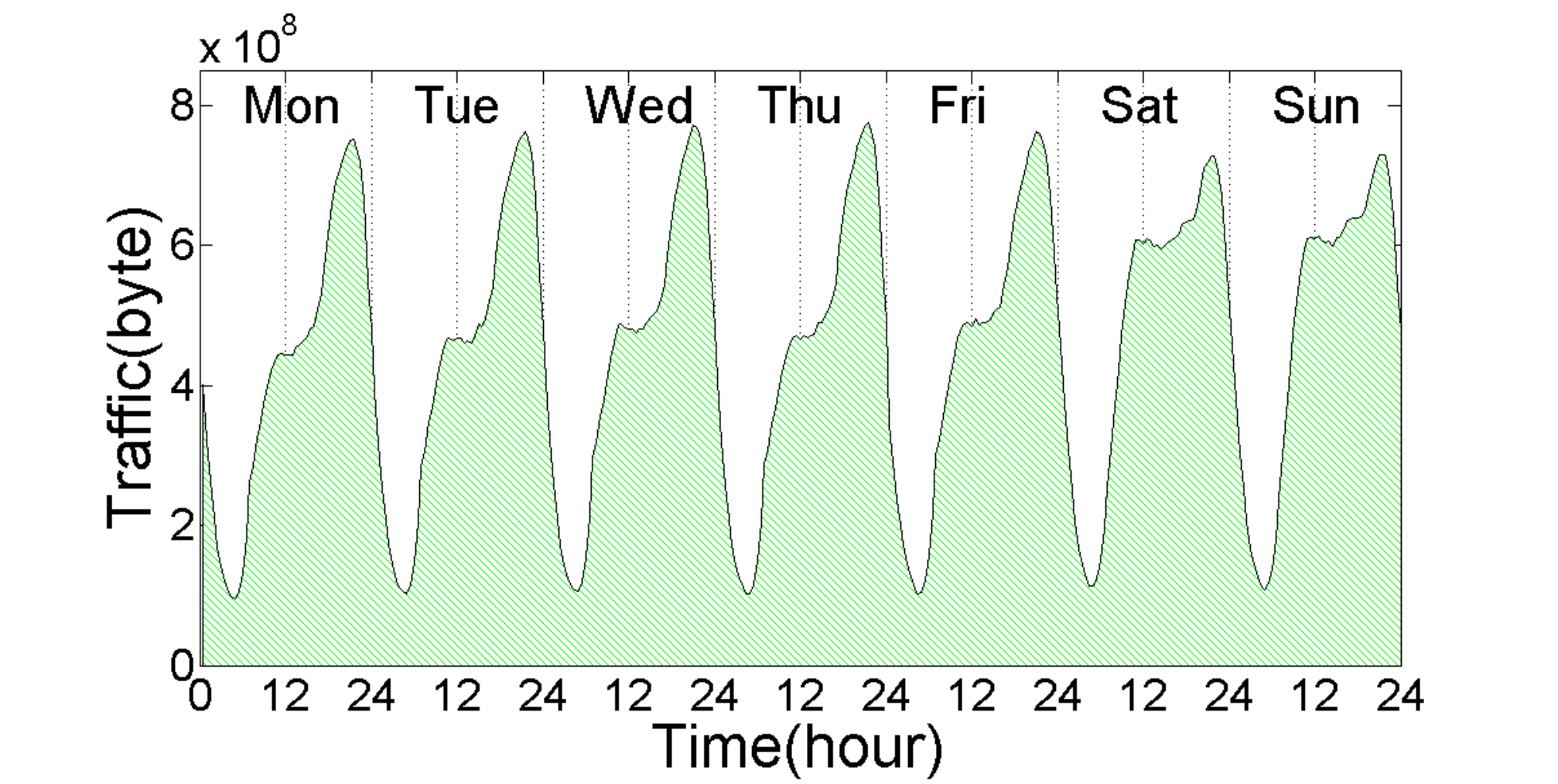}}
\subfigure[\#2:Transport area]{\includegraphics[width=.235\textwidth]{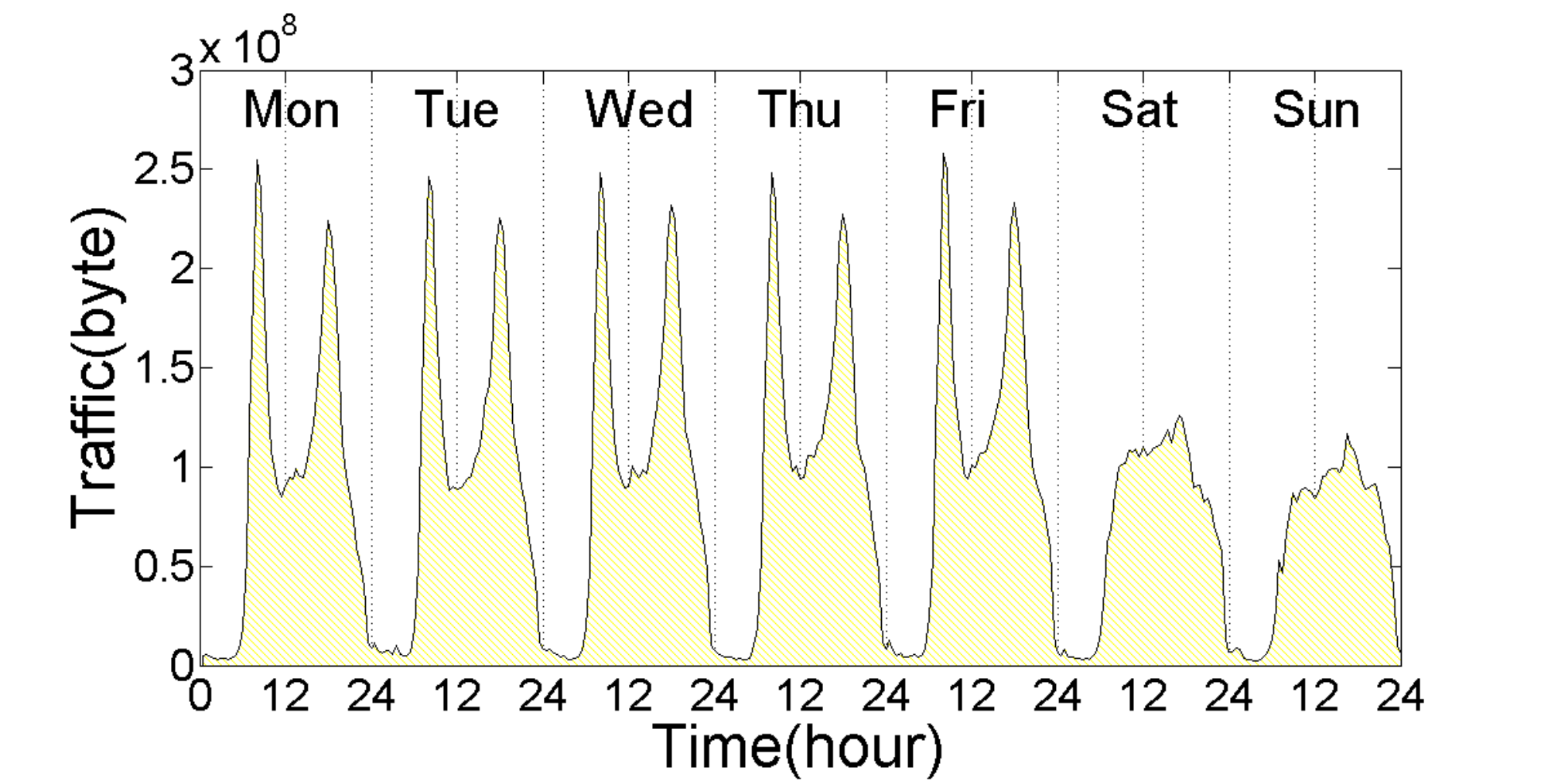}}
\subfigure[\#3:Office area]{\includegraphics[width=.235\textwidth]{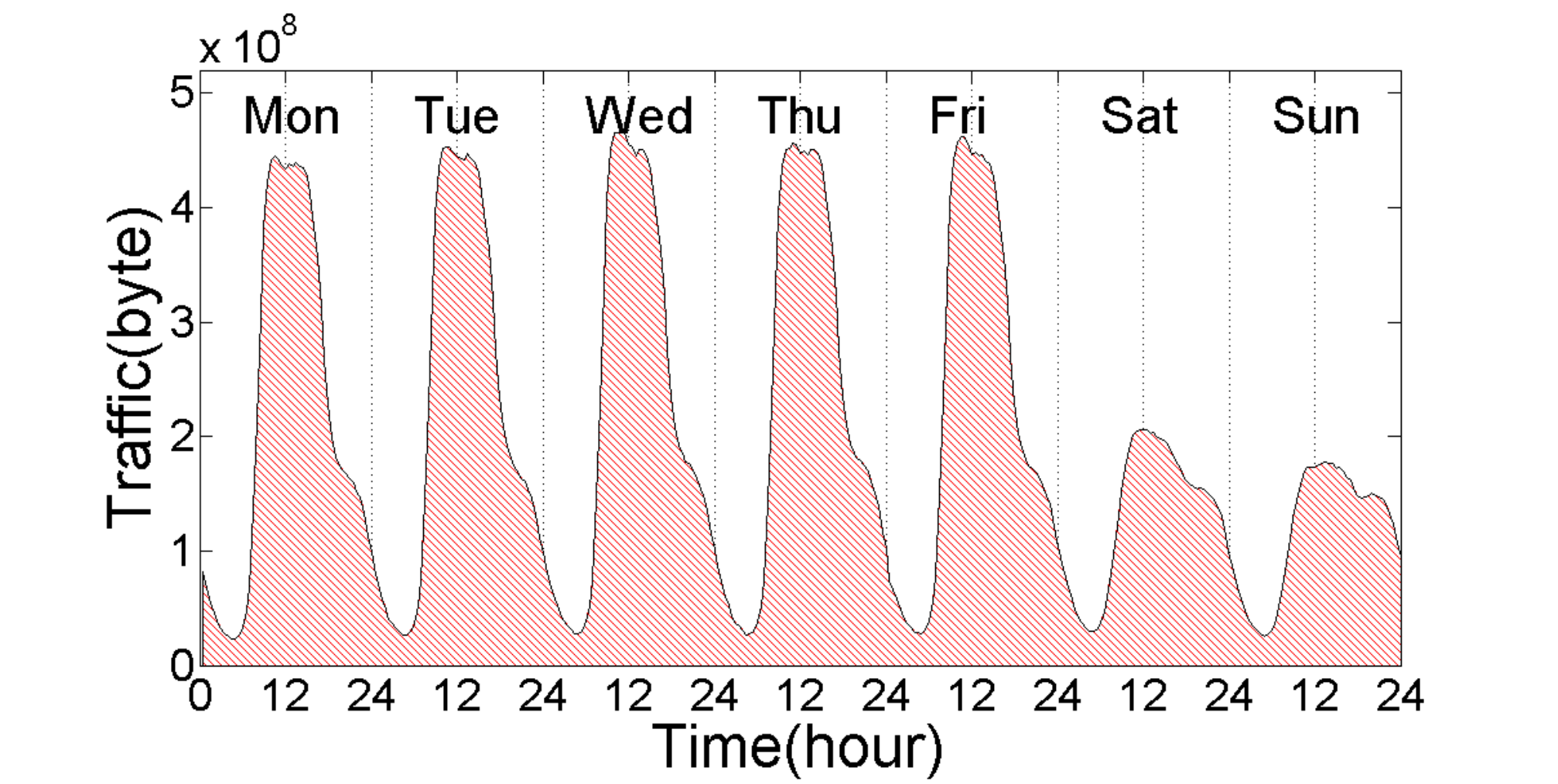}}
\subfigure[\#4:Entertainment area]{\includegraphics[width=.235\textwidth]{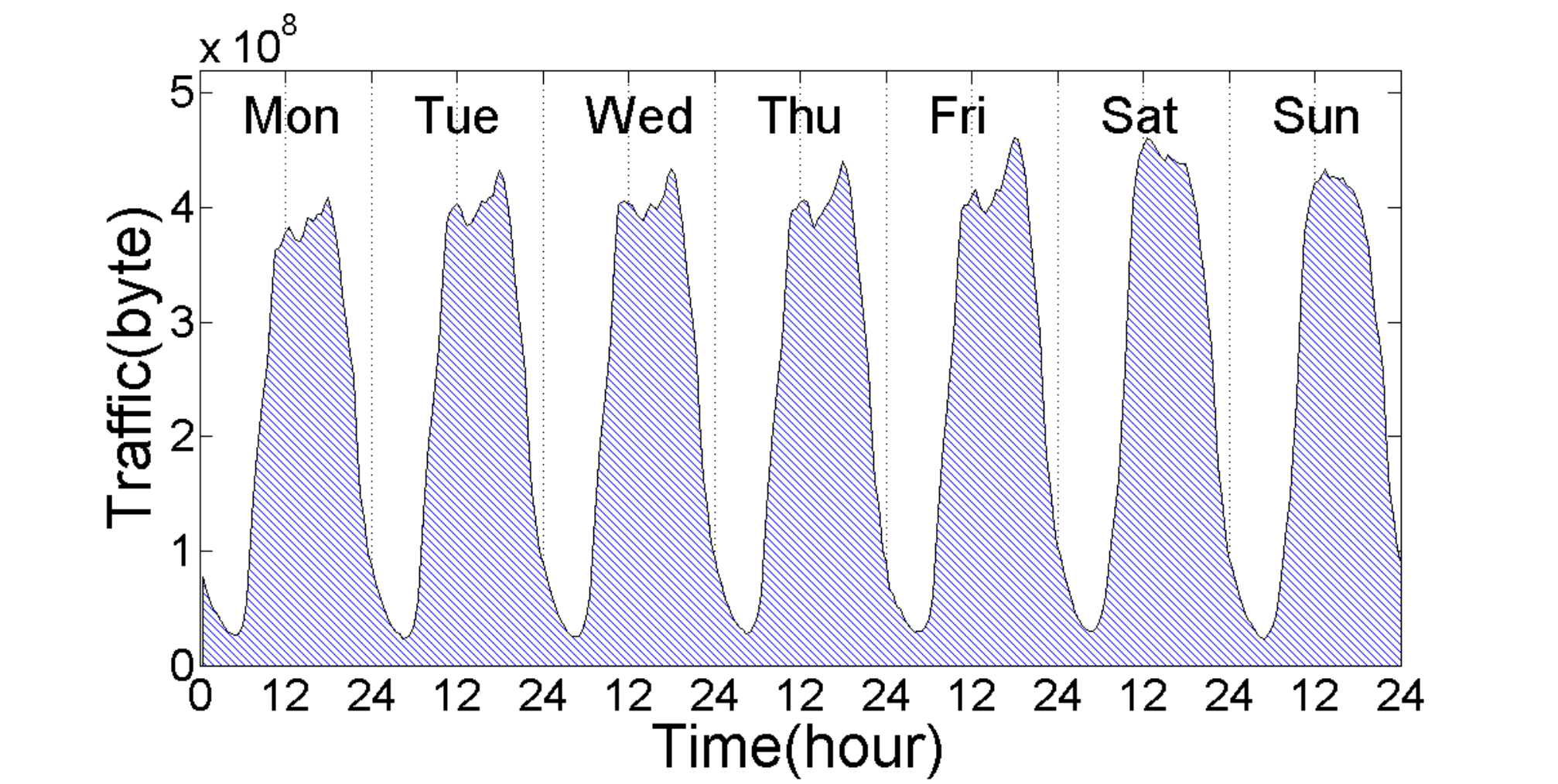}}
\subfigure[\#5:Comprehensive area]{\includegraphics[width=.235\textwidth]{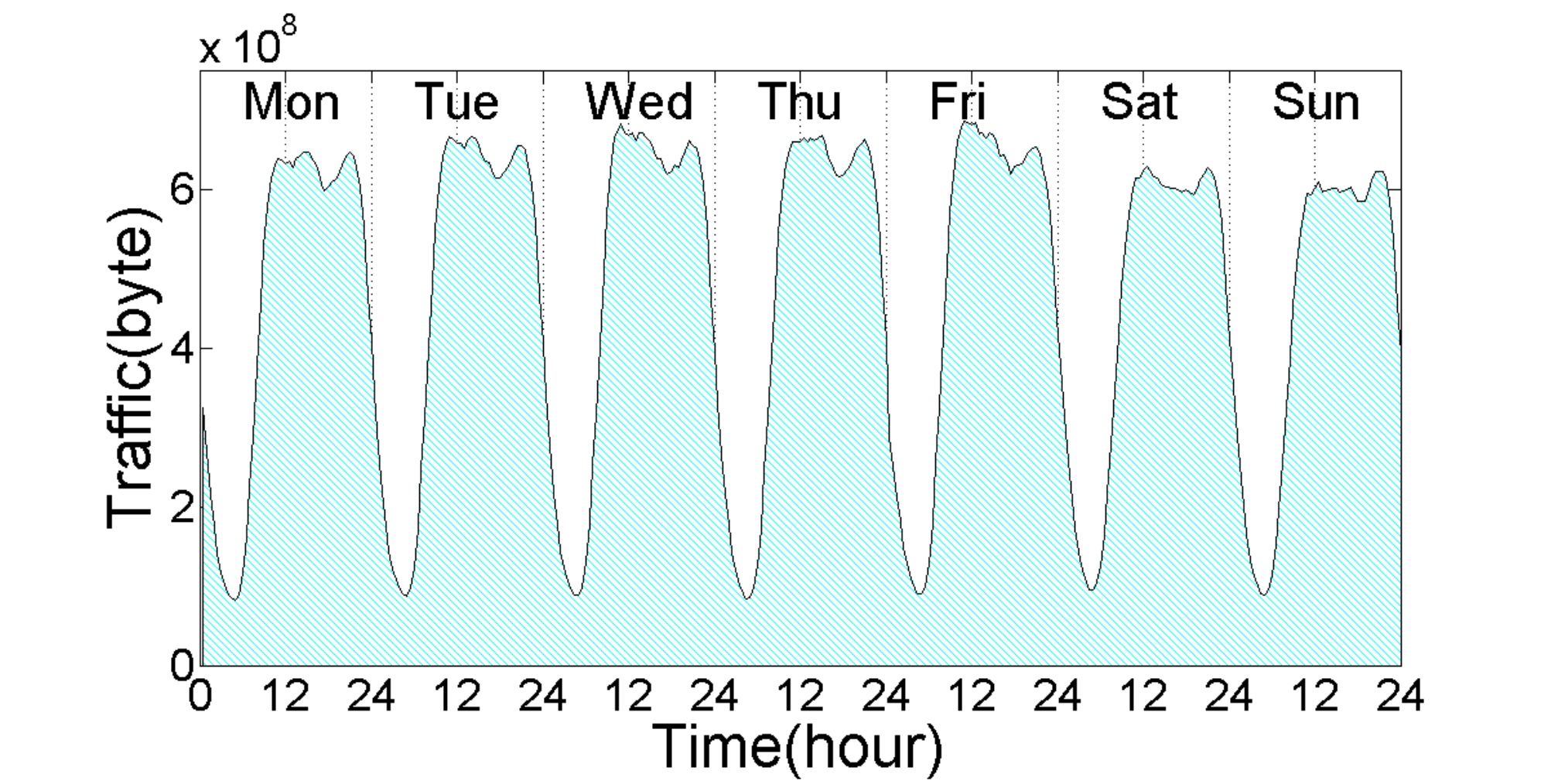}}
\caption{Patterns of the five identified clusters and CDF of clustering distance.} \label{fig:pattern}
\end{figure}

\mypara{Metric tuner:} As the number of traffic patterns is unknown, a key question is when the identifier should stop its clustering. In our system, we use Davies-Bouldin index \cite{maulik2002performance} to explicitly inform the identifier that the optimum number of patterns have been identified. Davies-Bouldin index is utilized because it measures both the separation of clusters and cohesion within clusters, which mathematically guarantees good clustering result. The mathematic formulation of Davies-Bouldin index is as follows,
\begin{equation*}
\begin{aligned}
& {\text{minimize}}
& & \frac{1}{R}\sum_{i=1}^{R}\max_{j=1, j\neq i}^{R}\frac{S_i+S_j}{M_{i,j}} , \\
& \text{subject to}
 & & M_{i,j} = ||A_i-A_j||_2 ,\\
 & & & S_i = \frac{1}{T_i}\sum_{k=1}^{T_i}||X_k-A_i||_2,
\end{aligned}
\end{equation*}
where the objective function is the Davies-Bouldin index, $X_i$ is the vectorized data of cellular tower $i$, $A_i$ is the centroid of each cluster, $R$ is the number of clusters and $T_i$ is the numbers of towers within the $i_{th}$ cluster. We minimize the Davies-Bouldin index by considering two factors --- the distance between clusters $M_{i,j}$ and $S_i$, which are the average distance from points to their cluster's centroid. When the minimum Davies-Bouldin index is obtained, the optimum number of patterns is identified. \PZ{The variation of DBI is shown in Figure~\ref{fig:pattern}(a), according to which we set the stop condition---threshold value at 16.33 to achieve optimal clustering result.}

Figure~\ref{fig:pattern} shows the five time-domain patterns identified by our system from the 9,600 cellular towers((c) to (g)) and each cluster's CDF of points' distance to its centroid(b). The five clusters differ in terms of the time where peak traffic appears as well as the amount of traffic experienced during weekday and weekends. Figure ~\ref{fig:pattern}(b) shows that the distance CDF curves of clusters are similar and all of them increase rapidly as distance increases. 80\% of points' distance to their clusters' centroid are less than 10, which implicates the clustering result is good. The percentage of each cluster's cell towers is shown in Table ~\ref{tab:clustercount}, which indicates the third cluster has most cell towers and second cluster the least.

\PZ{In conclusion, we implement a system that is able to identify the key traffic patterns among thousands of cellular towers in this subsection.} Since the five clusters are given by the hierarchical classifier, an interesting question to ask is what are the \PZ{geographical} locations where these five types of towers are deployed?

\begin{table}[b]
\begin{center}
\begin{tabular}{cccc}
\toprule
Cluster Index&Functional Regions&Percentage\\
\midrule
1&Resident&17.55\%\\
2&Transport&2.58\%\\
3&Office&45.72\%\\
4&Entertainment&9.35\%\\
5&Comprehensive&24.81\%\\
\bottomrule
\end{tabular}
\caption{Percentage of cell towers classified in each cluster.}\label{tab:clustercount}
\end{center}
\end{table}

\subsection{\PZ{Geographical} Context of Traffic Patterns}
To understand the \PZ{geographical} locations of cell towers of the five clusters, we first manually label typical towers in the five patterns with urban functional regions and then validate the labels of all towers in each pattern with ground truth.

\begin{figure} [t]
\begin{center}
\includegraphics*[width=8.8cm]{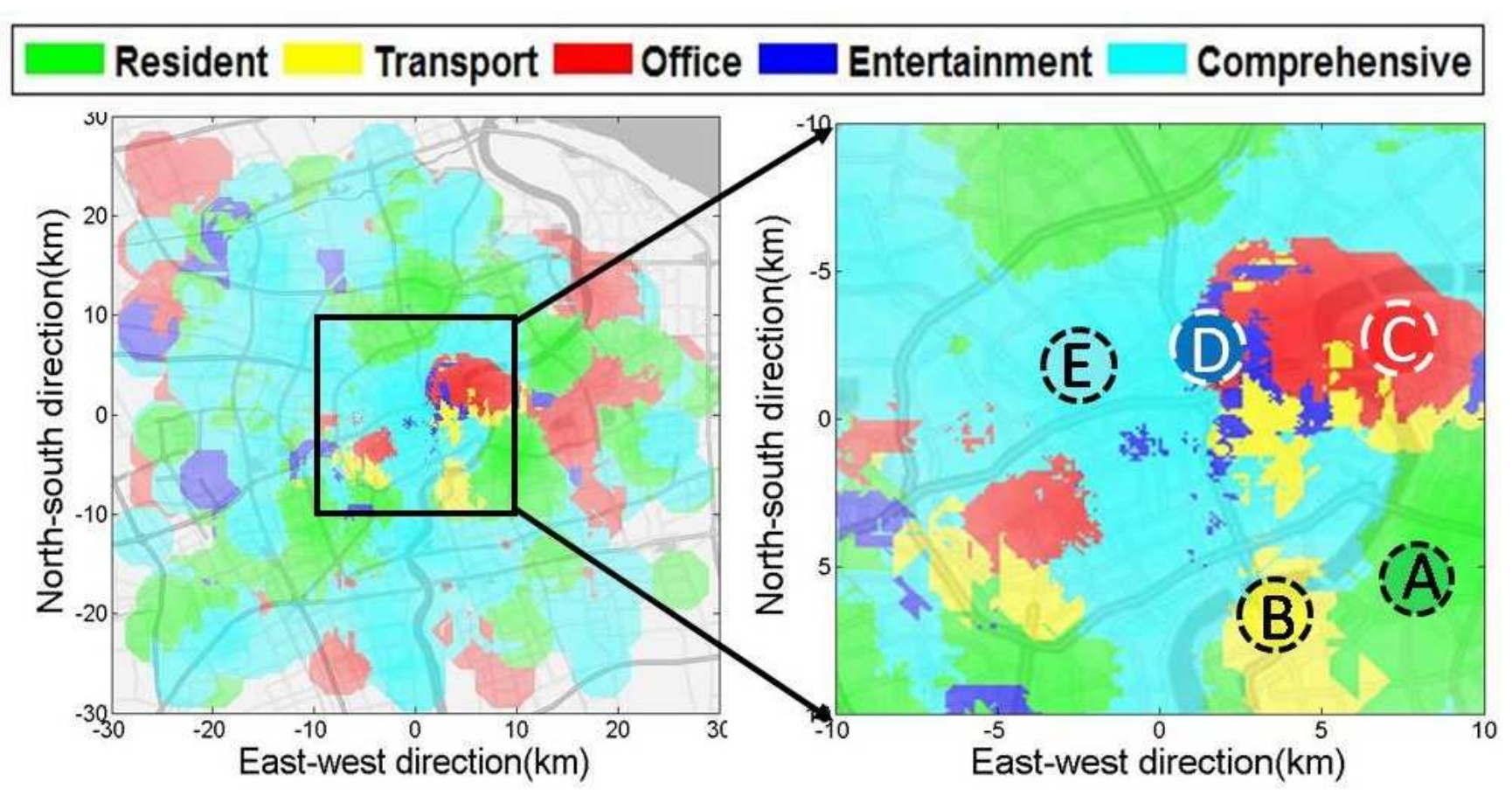}
\end{center}
\caption{\PZ{Geographical} distribution of base stations from the five identified patterns. } \label{fig:geophysical}

\end{figure}

\subsubsection{Label Patterns with Urban Functional Regions}
To understand the \PZ{geographical} context of traffic patterns, we label the five traffic patterns using urban functional regions. This process is nontrivial because given \PZ{thousands} of cellular towers, labelling cannot be done one by one manually. To address this challenge, we use a few human-labeled areas and combine with points of interests (POI) distribution to achieve accurate labelling. POI is a specific point location of a certain function such as restaurant and shopping mall. An area's POI distribution reflects its function. Therefore, studying POI distribution of one location can help us to accurately identify patterns' labels.
\PZ{The POI data we study is collected via APIs provided by Baidu Map introduced before. For calculating the POI distribution, we measure the number of four main types of POI, which are resident, transport, office and entertainment, within 200m of each cell towers.}
Figure~\ref{fig:geophysical} shows the \PZ{geographical} density map of towers in each cluster where deep color stands for higher density. Zooming in the urban area, for each cluster we pick the point with the highest tower density and calculate their POI distribution as summarized in Table~\ref{tab:HeatPOI}. Then, we infer the urban function region of each cluster by checking the \PZ{geographical} location information in Figure~\ref{fig:geophysical} and POI distribution in Table~\ref{tab:HeatPOI}. We obtain the following \PZ{geographical} labels for the five clusters.

\mypara{Resident area:}
Figure~\ref{fig:geophysical} shows that the towers in this cluster (green color) are mainly distributed on the surrounding areas of the city. In addition, the highest density point, \textbf{\emph{A}}, is located in a large resident neighborhood. Table~\ref{tab:HeatPOI} also shows that the number of residential points in \textbf{\emph{A}} is more than others. Therefore,we label the area covered by this cluster's cell towers as residential area.

\mypara{Transport area:}
 In Figure~\ref{fig:geophysical}, the second cluster's highest density point \textbf{\emph{B}} is close to an area with three subway stations and one overpass. In addition, Table ~\ref{tab:HeatPOI} shows that  around location \textbf{\emph{B}} the number of transport POI is higher than the rest even though its absolute number is small. Therefore, we label this cluster as transport area.

\mypara{Office area:}
Figure~\ref{fig:geophysical} shows that the highest density point \textbf{\emph{C}} is a well-known business district in Shanghai. This location mark is also verified by the third row of Table~\ref{tab:HeatPOI} where the number of office POI is dominant for the area 200m from \textbf{\emph{C}}. As a result, we label this cluster as office area.

\mypara{Entertainment area:}
The highest density point \textbf{\emph{D}} in Figure~\ref{fig:geophysical} is a large shopping mall and entertainment park in Shanghai. Table~\ref{tab:HeatPOI} also shows that its number of entertainment POI is more than the rest. Therefore, we label this cluster as entertainment area.

\mypara{Comprehensive area:}
Figure~\ref{fig:geophysical} shows the tower density map of the last cluster, where we observe uniform distribution of towers across the city. In addition, the highest density point, \textbf{\emph{E}}, is a comprehensive area, which includes all kinds of urban functions, including residential area, offices, etc. The POI distribution of point E does not suggest obvious land mark either. Therefore, it is labeled as comprehensive area.

\begin{table}[t]
\begin{center}
\caption{Distribution of POI at Chosen Point.}\label{tab:HeatPOI}
\begin{tabular}{|p{1.2cm}<{\centering}|p{1cm}<{\centering}p{1cm}<{\centering}p{1cm}<{\centering}p{1cm}<{\centering}|}\hline
\multirow{2}{*}{Point}  & \multicolumn{4}{c|}{Points of Interest} \\ \cline{2-5}
& \multicolumn{1}{c}{Resident} & \multicolumn{1}{c}{Transport} & \multicolumn{1}{c}{Office$\ $ $\ $} & \multicolumn{1}{c|}{Entertain}\\ \hline
\textbf{\emph{A}}   &\cellcolor[rgb]{.7,.9,.9}195 &0 &19  & 51\\
\textbf{\emph{B}}   &68 &\cellcolor[rgb]{.7,.9,.9}2 &56 &36\\
\textbf{\emph{C}}   &151 &1 &\cellcolor[rgb]{.7,.9,.9}1016 &157\\
\textbf{\emph{D}}   &16 &0 &108 &\cellcolor[rgb]{.7,.9,.9}2165\\
\textbf{\emph{E}}   &59 &0 &179 &26\\\hline
\end{tabular}
\end{center}

\end{table}

\subsubsection{Validate the Labels }
In this section, we validate the labels of the five patterns in both micro and macro scale. Our labels are obtained by checking the \PZ{geographical} locations of a few towers in each cluster and verifying with the corresponding POI distribution. However, the correctness of labelling across all 9,600 cellular towers remains unknown. Therefore, we perform further analysis to validate our labels \PZ{with POI data} from  micro and macro two perspectives.
\begin{figure}[t]
\centering
\subfigure[Area A]{\includegraphics[width=.235\textwidth]{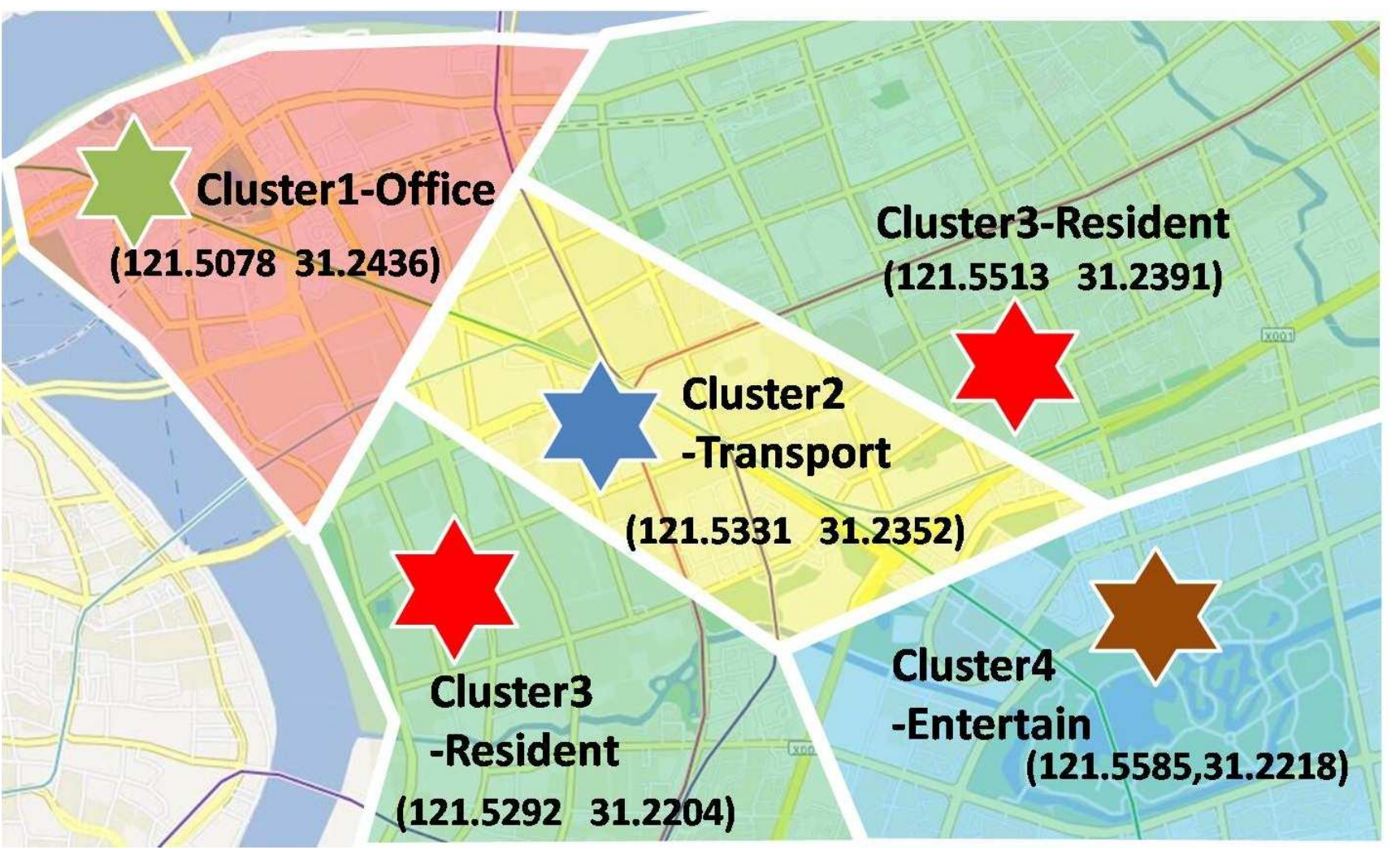}}
\subfigure[Area B]{\includegraphics[width=.235\textwidth]{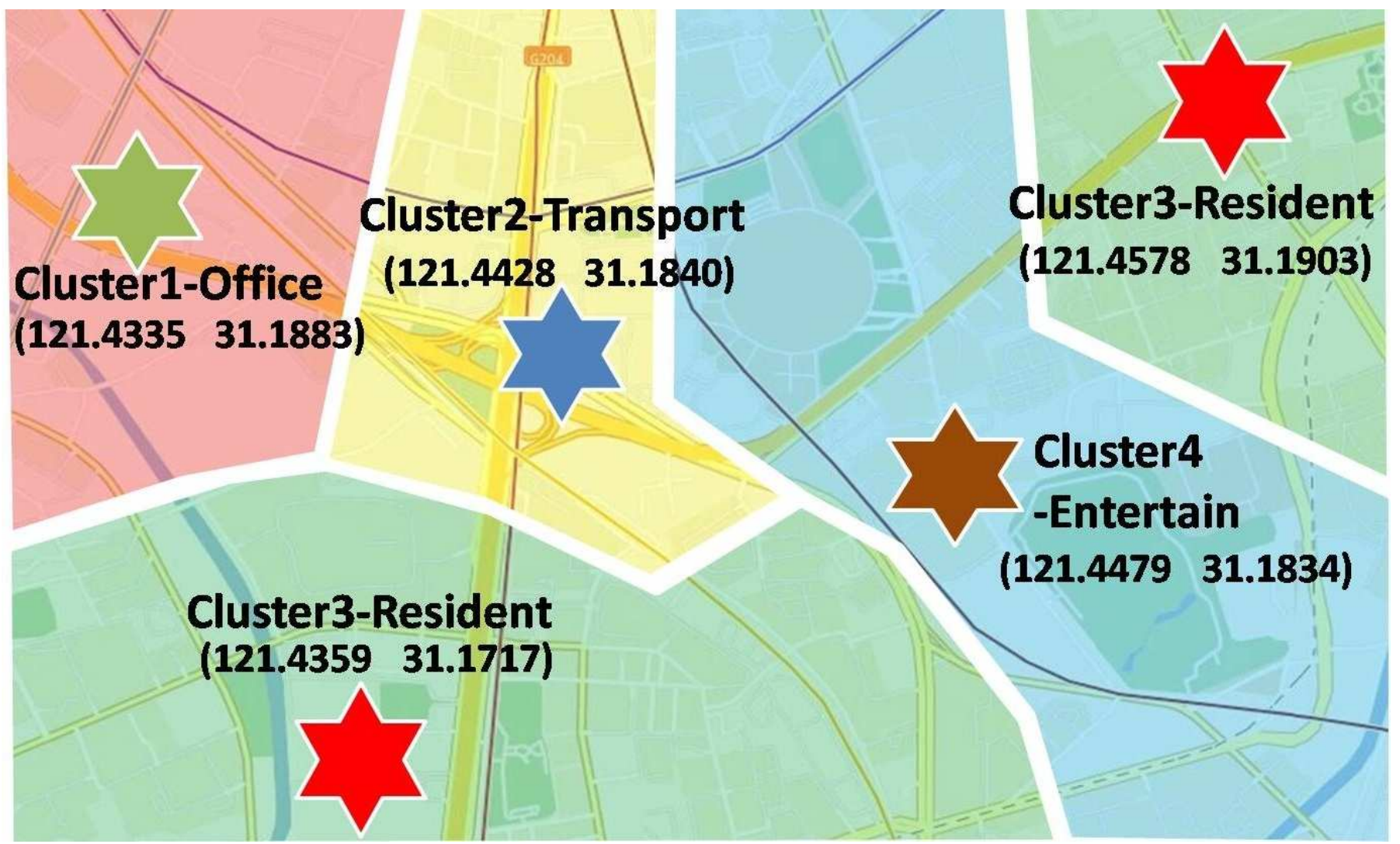}}
\caption{Two case studies for validating the \PZ{geographical} context of the five identified patterns.} \label{fig:casestudy}

\end{figure}

\mypara{Validate with case study:}
To validate our labels in micro scale, we randomly choose two areas shown in Figure~\ref{fig:casestudy}. According to the POI data, we first color different functional regions in the area with different colors. Green represents residential area, yellow represents transport area, red represents office area, and blue represents entertainment area. After that, we investigate the labels of cell towers locating in the area. Observing both Figure~\ref{fig:casestudy}(a) and (b), we find that the labels attached to the cell towers exactly match with the functional regions, which justifies our labels' correctness.

\begin{figure} [t]
\begin{center}
\includegraphics*[width=8.8cm]{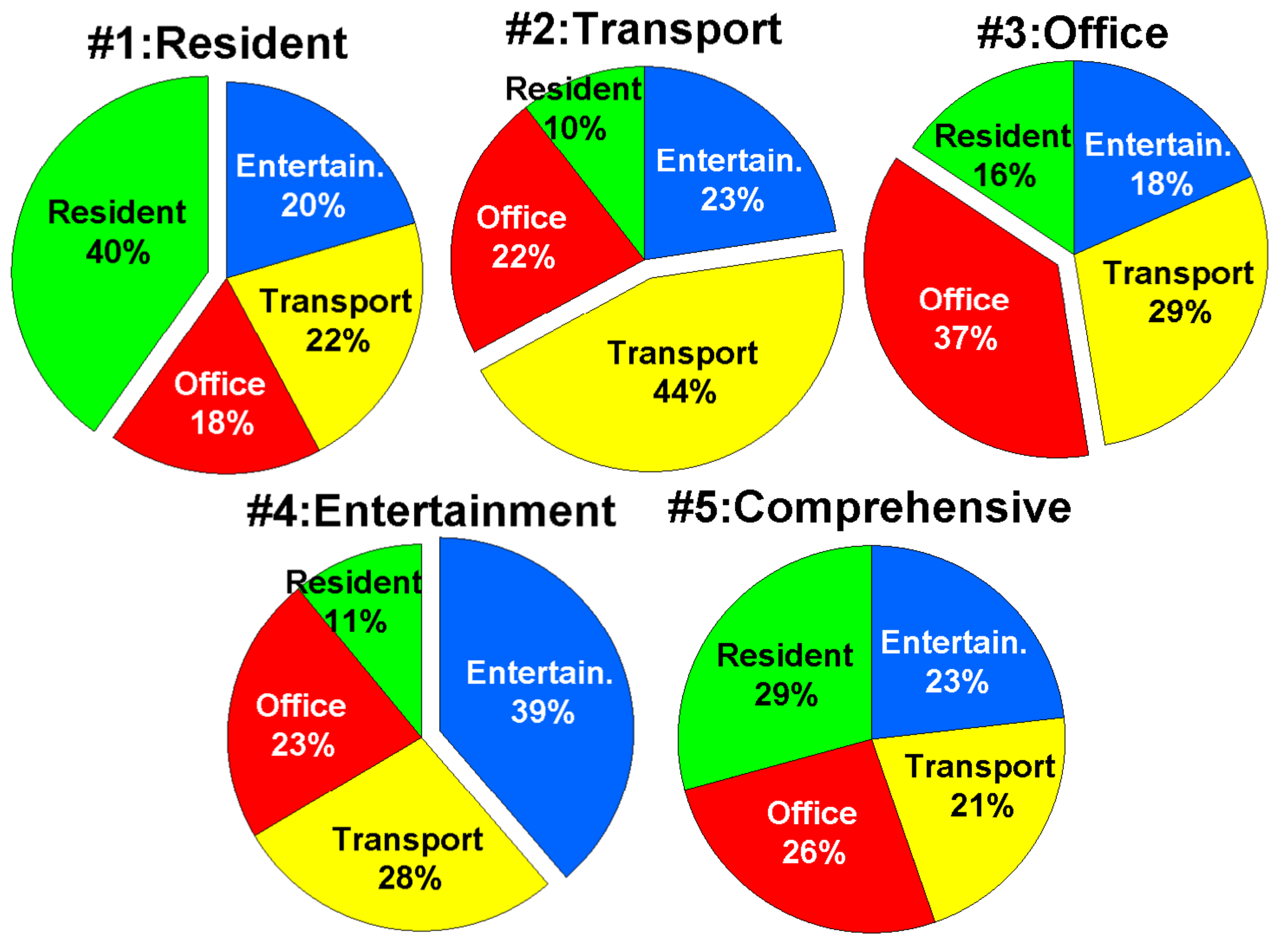}
\end{center}
\caption{Pie chart of averaged normalized points of interest of five clusters. }

\label{fig:pie}
\end{figure}

\mypara{Validate with 9,600 towers' POI:}
To validate our labels in macro scale, we perform further analysis on all 9,600 towers' POI. However, different types POI vary in magnitude significantly because of their different nature. To eliminate this interference, we first perform min-max normalization on each type's POI and then average them by clusters, which is summarized in Table~\ref{tab:NormalPOI}.
The maximum of each row and column is marked with color, which shows the dominant urban function in each cluster. Figure~\ref{fig:pie} explicitly shows each POI's percentage in five clusters. According to Table~\ref{tab:NormalPOI} and Figure~\ref{fig:pie}, transport type POI dominates the region labeled as transport area, with 44\% of this area's POI, while entertainment area is dominated by entertainment type POI for 39\%. These measurements validate the labels obtained from the sampled towers of each cluster.

\begin{table}[t]
\begin{center}
\caption{Averaged normalized points of interest of five clusters.}\label{tab:NormalPOI}
\begin{tabular}{|p{1.2cm}<{\centering}|p{1cm}<{\centering}p{1cm}<{\centering}p{1cm}<{\centering}p{1cm}<{\centering}|}\hline
\multirow{2}{*}{Cluster}  & \multicolumn{4}{c|}{Points of Interest} \\ \cline{2-5}
 & \multicolumn{1}{c}{Resident} & \multicolumn{1}{c}{Transport} & \multicolumn{1}{c}{Office $\ $ $\ $} & \multicolumn{1}{c|}{Entertain}\\ \hline
\#1    &\cellcolor[rgb]{.7,.9,.9}0.0528 &0.0285 & 0.0232 &0.0269\\
\#2   &0.0473 &\cellcolor[rgb]{.7,.9,.9}0.2000 & 0.1012&0.1020\\
\#3   &0.0439 &0.0813 & \cellcolor[rgb]{.7,.9,.9}0.1034 & 0.0515\\
\#4   &0.0474 &0.1201  &0.0976 &\cellcolor[rgb]{.7,.9,.9}0.1674 \\
\#5   &0.0508 &0.0373 & 0.0453 &0.0403\\\hline
\end{tabular}
\end{center}

\end{table}

\PZ{To conclude, in this subsection we verify our identified key traffic patterns as well as establish their relationships with urban functional regions.}

\section{Understanding Modeled Traffic Patterns: Time Domain Aspect}
Understanding the hidden physical meanings of traffic patterns is important for exploiting them to solve practical problems, such as traffic load balancing or land usage identification. Although we have identified key traffic patterns and linked them to corresponding urban functional regions, we still have little knowledge of the hidden physical meaning of these patterns. In this section, we conduct an analysis to reveal the time and \PZ{geographical} characteristics of modeled traffic patterns.

\begin{figure}[t]
\centering
\subfigure[Weekday-weekend traffic amount ratio.]{\includegraphics[width=.235\textwidth]{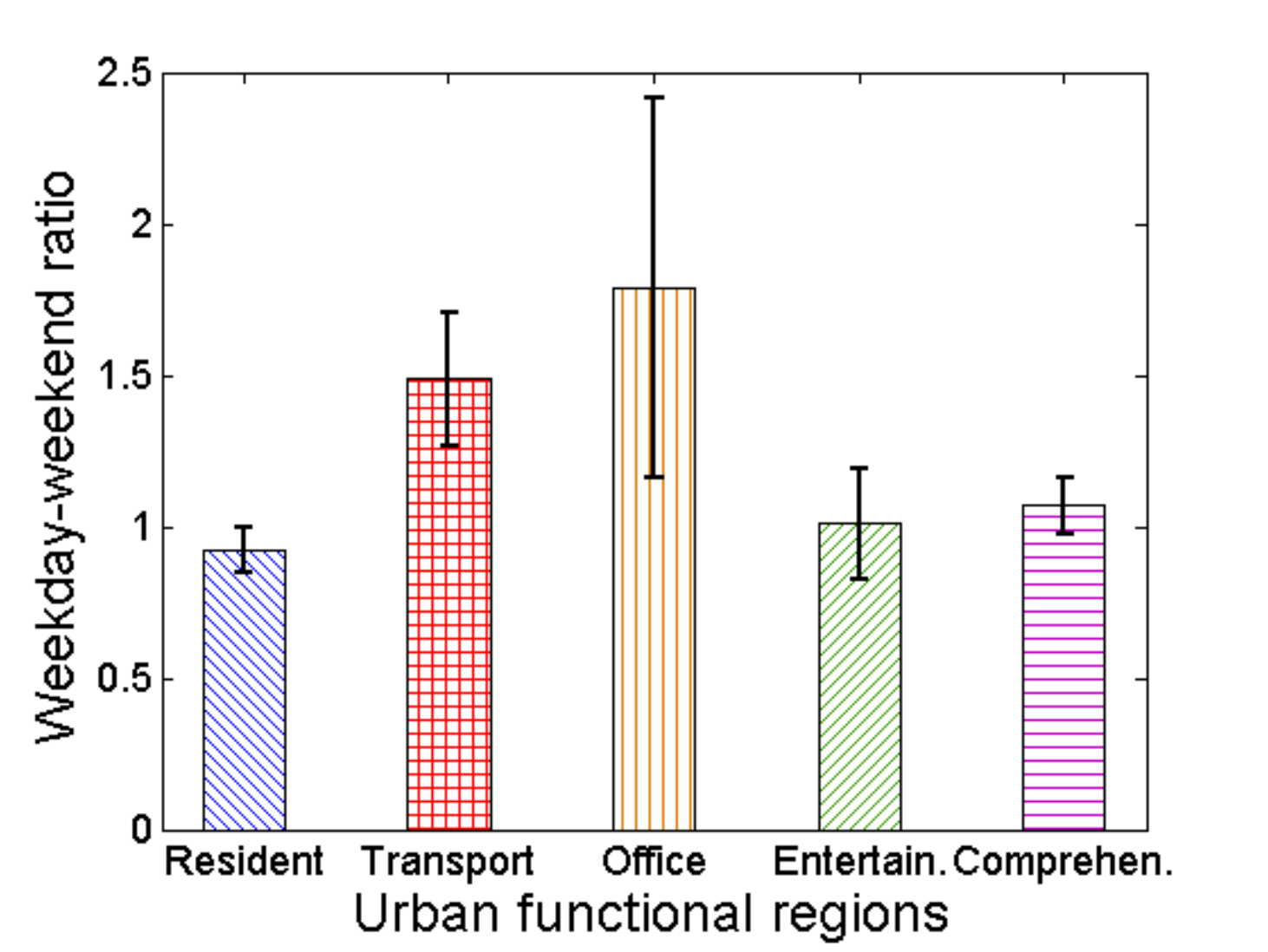}}
\subfigure[Weekday and weekend's peak-valley ratio]{\includegraphics[width=.235\textwidth]{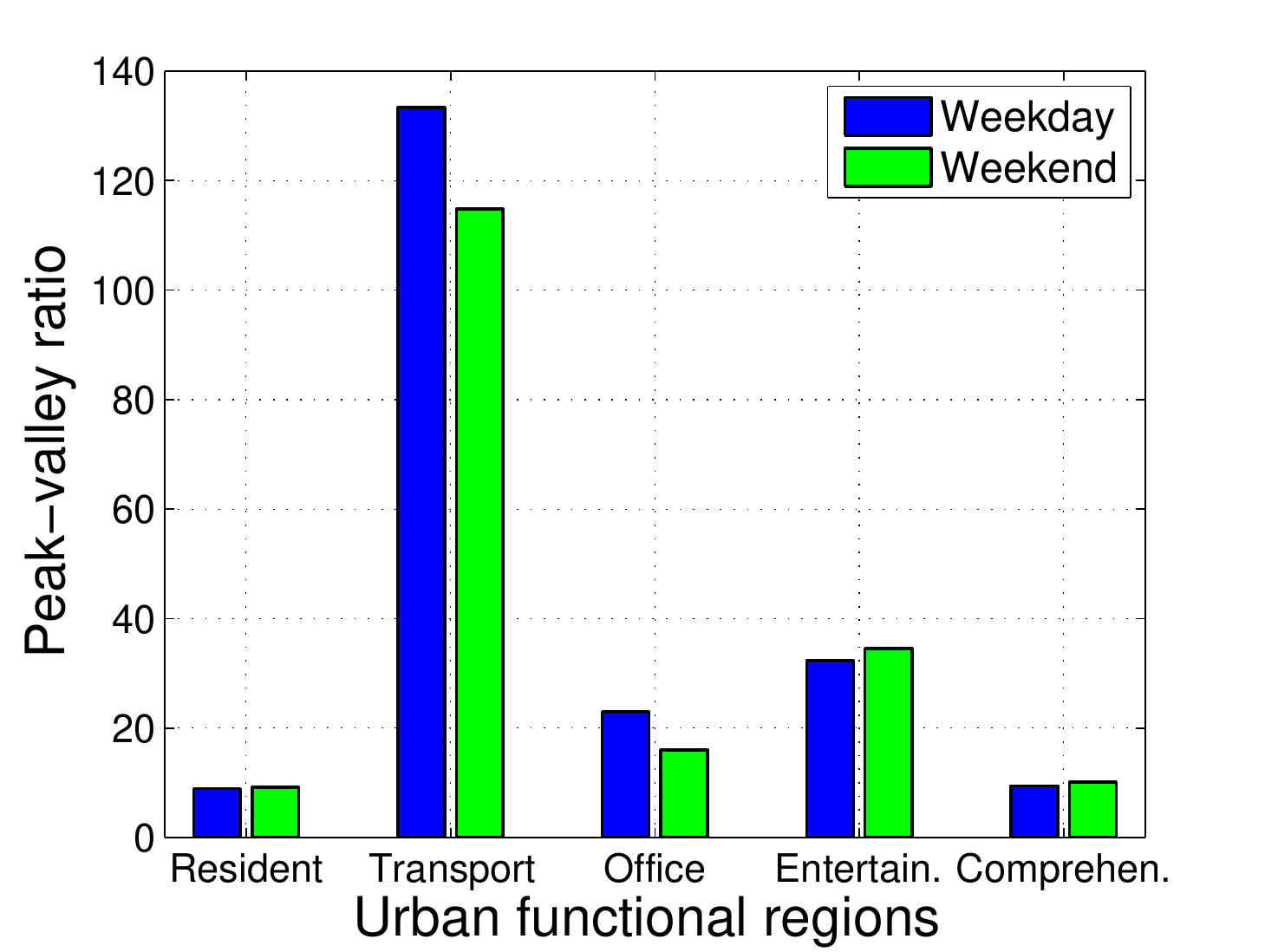}}
\caption{Time-domain characteristics of the five identified patterns.} \label{fig:timecharacter}

\end{figure}

We start from quantifying the distinct time-domain characteristics of each pattern and find out the interrelationships between these patterns. These time-domain characteristics have not been quantified and the interrelationships between these traffic patterns remain unknown, which are two obstacles preventing us from further understanding these traffic patterns. Therefore, in order to understand the hidden meaning of modeled patterns, we conduct the following studies.

%

\subsection{Quantify Time-Domain Characteristics}
It is obvious that traffic patterns of different urban functional regions possess different characteristics in time-domain. In this subsection, we dedicate to quantify these characteristics and provide insights of traffic behaviours in different urban functional regions.

\mypara{Weekday-Weekend traffic amount ratio:} Observing Figure~\ref{fig:pattern}, traffic amount during \PZ{weekday is significantly different} from weekend in transport area and office area. We quantify this characteristic by computing the ratio between weekday's traffic amount and weekend's, which is presented in Figure~\ref{fig:timecharacter}(a). According to Figure~\ref{fig:timecharacter}(a), one day's traffic amount in resident area, entertainment area and comprehensive area is almost identical between weekday and weekend. However, weekday-weekend traffic amount ratio in transport area is 1.49 and the ratio in office area is 1.79, which suggests weekday's traffic amount of those two regions is much more than weekend. This phenomenon makes sense because \PZ{people typically go to work} in weekday while they do not in weekend.

\begin{table*}[bth]
\tiny
\begin{center}
    \begin{tabular}{|c|c|c|c|c|c|c|c|c|c|c|}
    \hline
    \multirow{2}{*}{\backslashbox{Features}{Regions}}  & \multicolumn{2}{c|}{resident area} & \multicolumn{2}{c|}{transport area} & \multicolumn{2}{c|}{office area} & \multicolumn{2}{c|}{entertainment area} & \multicolumn{2}{c|}{comprehensive area} \\ \cline{2-11}
    ~                & weekday            & weekend & weekday       & weekend & weekday      & weekend & weekday      & weekend & weekday            & weekend \\ \hline
    maximum traffic  & $7.77 \times 10^8$ & $7.99 \times 10^8$ & $2.76 \times 10^8$ & $1.55 \times 10^8$ & $4.69 \times 10^8$ & $2.78 \times 10^8$ & $4.55 \times 10^8$ & $4.90 \times 10^8$ & $7.36 \times 10^8$ & $7.38 \times 10^8$ \\ \hline
    minimum traffic  & $8.70 \times 10^7$ & $8.71 \times 10^7$ & $2.07 \times 10^6$ & $1.35 \times 10^6$ & $2.04 \times 10^7$ & $1.74 \times 10^7$ & $ 1.41 \times 10^7$ & $1.42 \times10^7$ & $7.77 \times 10^7$ & $7.29 \times 10^7$ \\ \hline
    peak-valley ratio    & 8.93               & 9.17    & 133.33          & 114.81    & 22.99       & 15.98   & 32.27        & 34.51    & 9.47              & 10.12   \\ \hline
    \end{tabular}
    \caption {Peak-valley features}
    \label{tab:maxmin}
\end{center}

\end{table*}

\begin{table*}[bth]
\tiny
\begin{center}
    \begin{tabular}{|c|c|c|c|c|c|c|c|c|c|c|}
    \hline
    \multirow{2}{*}{\backslashbox{Features}{Regions}}  & \multicolumn{2}{c|}{resident area} & \multicolumn{2}{c|}{transport area} & \multicolumn{2}{c|}{office area} & \multicolumn{2}{c|}{entertainment area} & \multicolumn{2}{c|}{comprehensive area} \\ \cline{2-11}
    ~                & weekday            & weekend & weekday       & weekend & weekday      & weekend & weekday      & weekend & weekday            & weekend \\ \hline
    time of peak    &  21:30          & 21:30  & 8:00 18:00          &     & 10:30       &  12:00    & 18:00        & 12:30   &              &    \\ \hline
    time of valley  &  5:00          & 5:00  & 4:00           &  4:30   & 5:00       &  5:00    & 5:00        & 5:0   &     5:00         & 5:00   \\ \hline
    \end{tabular}
    \caption {Time of traffic peak and valley}
    \label{tab:timepeak}
\end{center}
\end{table*}

\mypara{Peak-valley features:}Observing Figure~\ref{fig:pattern}, all traffic patterns experience periodic peaks and valleys. However, the traffic patterns are significantly different in peak value, valley value and peak-valley ratio. We quantify these characteristics and summarize them in Table~\ref{tab:maxmin}. According to Table~\ref{tab:maxmin}, in transport area and office area weekend's  maximum traffic and minimum traffic is much less than weekday, which is consistent with last paragraph's finding. What's more, the transport's peak-valley ratio is much higher than other regions, which is explicitly presented in Figure~\ref{fig:timecharacter}(b). However, transport area's maximum traffic is less than other regions both in weekday and weekend. It suggests that transport area has the least traffic amount and the largest peak-valley traffic difference, while resident area and comprehensive area are the opposite.

\mypara{Time of traffic peak and valley:} Different urban functional regions' traffic patterns differ not only in peak volume, but also in peak time. We quantify this characteristic and present it in Table~\ref{tab:timepeak}. We leave the blank unfilled, if there is not a periodic peak or valley. Observing Table~\ref{tab:timepeak}, we find that traffic valley always takes place in 4:00$\sim$5:00. In weekday, transport area has two peaks in 8:00 and 18:00, which are probably caused by rush hour. In entertainment area, weekday's traffic peak time is 18:00 while weekend's traffic peak time is 12:30. It suggests that people go for entertainment later in weekday because of work.

\PZ{To conclude, we quantify the time-domain characteristics of each identified traffic pattern, which paves the way towards a deep understanding of cellular traffic patterns.}

\begin{figure} [t]
\begin{center}
\includegraphics*[width=8.8cm]{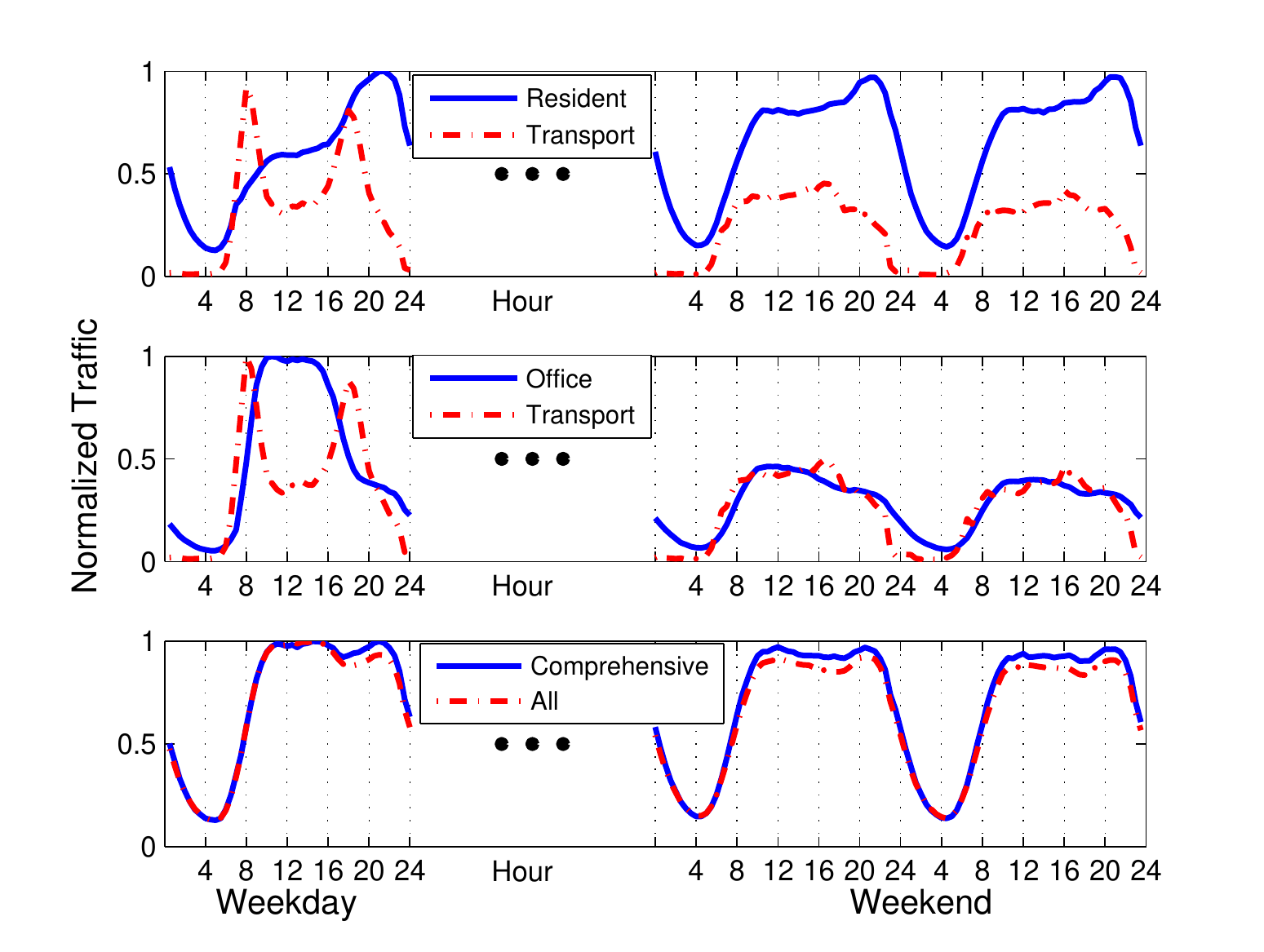}
\end{center}
\caption{Understanding the interrelationships between traffic patterns. }
\label{fig:interrelationship}
\end{figure}

\subsection{Interrelationships Between Traffic\\ Patterns}
We compare the interrelationships between normalized modeled traffic patterns in Figure~\ref{fig:interrelationship}. The first row of Figure~\ref{fig:interrelationship} compares the modeled traffic patterns of residential areas and transport hot spots. The peak of residential area is about 3 hours later than the second peak of transport, and the slope of these two peaks is almost identical. In addition, when we compare traffic patterns of transport hot spots and business district shown in Figure~\ref{fig:interrelationship}, we find that the peak in business district takes place in the time period between the two peaks of transport hot spots. Both observations suggest that the traffic patterns in these three areas are related. These three traffic patterns probably depict the daily routine of working populations, for them rush through heavy traffic area to work in morning and rush back home in evening.

In the third row of Figure~\ref{fig:interrelationship}, blue line stands for the traffic pattern in comprehensive area, and red line stands for the average traffic pattern of all cell towers. In fact, we find that these two patterns are of great similarity, which suggests that comprehensive area really is a mixture of other four kinds of functional areas.
\PZ{In conclusion, we analyze the interrelationships between the traffic patterns of different urban functional regions, which provides insightful understanding.}

\section{Frequency-domain Representation for Traffic Modeling}

In this section, we conduct frequency-domain analysis.
Such frequency-domain analysis is motivated by observing
the inherent time-domain periodicity of traffic and
the disadvantages of pure time-domain traffic analysis, where time-domain traffic identification is not easy, especially when cellular towers are deployed in the comprehensive areas with couples of behaviors.
For example, we know that traffic of cellular tower in the office area reaches the valley in weekends, and traffic of cellular tower in transport area has two peaks in one day, but for an arbitrary cellular tower which has both characteristics,
we do not know which of the two will predominate. On the other hand, in frequency domain, we can quantify these characteristics by using the amplitude and phase of frequency corresponding to one day and one week. Thus, we can grasp the key points and compare the strength of different characteristics of traffic for one cellular tower,
which is not intuitive in time domain.
Here, a natural question to ask is what are the most discriminating and essential features to present traffic patterns of cellular towers. Motivated by answering this question, we conduct frequency domain analysis on the five extracted patterns and reveal several important discoveries.

\subsection{Frequency Transform}

\begin{figure}[t]
\centering
\subfigure[]{\includegraphics[width=.49\textwidth]{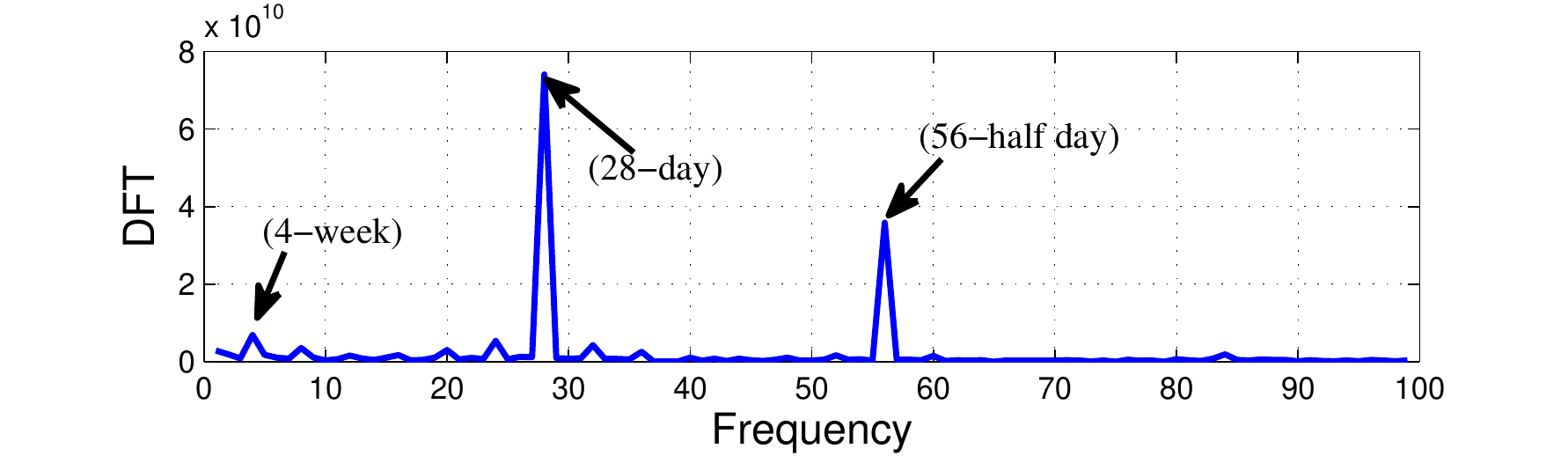}}
\subfigure[]{\includegraphics[width=.49\textwidth]{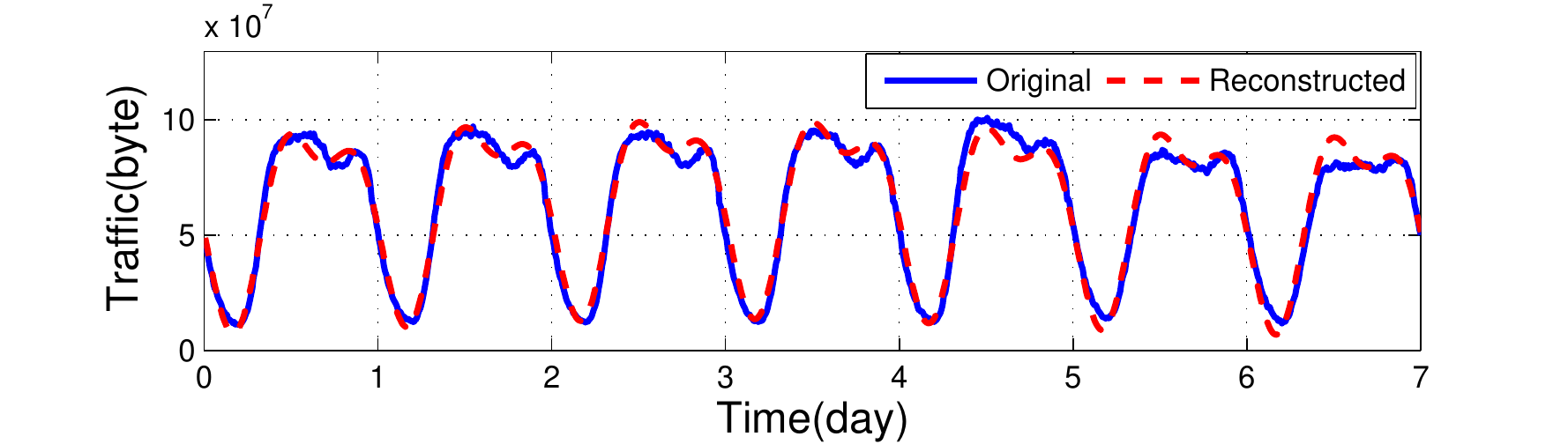}}
\caption{\WHD{Time-domain traffic reconstructed by the three principal frequency-domain components (k=4, 28, 56).}}\label{fig:fexa}
\end{figure}

In order to analyze the strong periodicity existing in time domain,
we first carry out discrete fourier transform (DFT) on the time-domain traffic vector \WHD{$X\!=\!(x[1],...,x[N])^T$}.
\WHD{$X$ can be either the time-domain traffic vector of one cellular tower, $i.e.$, $X_j$ for cellular tower $j$, or the aggregate traffic vector of a cluster, $i.e.$, $\sum_{j\in C}X_j$ for the cluster $C$.
The process} can be formulated as the following:
$$
\setlength{\abovedisplayskip}{0pt}
\setlength{\belowdisplayskip}{0pt}
\hat{X}[k]=\sum_{n=1}^{N}x[n]e^{-2\pi ikn/N},
$$
where $N$ is the number of traffic samples, that is \WHD{28 days' 10-minutes segmentation, $i.e.$, 4032} as discussed before in our analysis.
$\hat{X}[k]$ is the frequency spectrum of time-domain traffic $X$. \WHD{Figure~\ref{fig:fexa}(a)} shows the DFT
\WHD{of the aggregate traffic of all cellular towers}, where three peaks are observed, $i.e.$ $k=$4, 28, 56.
\WHD{Since the duration of our series is 4 weeks, the 4$th$ point is corresponding to time-domain periodic patterns of one week.
Similarly, the 28$th$ and 56$th$ points stand for the time-domain periodic patterns of one day and half a day, respectively.} The absolute values of the three components are much higher than the rest of points, which suggests that most information of the time-domain traffic could be retained by the three components.
Motivated by this hint, we use the three components for presenting the time-domain traffic. To evaluate the information loss of ignoring the rest of frequency components, we reconstruct the time-domain traffic using the three main frequency components, which is expressed as follows:
\begin{eqnarray*}
\setlength{\abovedisplayskip}{0pt}
\setlength{\belowdisplayskip}{0pt}
\ \ \ \
\begin{cases}
\hat{X}^r[k]\!=\!\!
\begin{cases}
\hat{X}[k],&\!\!  \text{if $k=$0, 4, 28, 56, $N$-4, $N$-28, $N$-56,} \\
0,&\!\!  \text{otherwise,}
\end{cases}\\
\\
x^r[n]=\frac{1}{N}\sum_{k=0}^{N-1}\hat{X}^r[k]e^{2\pi ikn/N},
\end{cases}
\end{eqnarray*}

where $x^r[n]$ is the reconstructed time-domain traffic.
\WHD{The reconstructed time-domain traffic of the aggregate traffic of all cellular towers
is also shown in Figure~\ref{fig:fexa}(b).}
From the result, we can observe that the reconstructed curve is very close to the original curve.
Specifically, the lost energy, $\sum_{n=1}^N x^r[n]^2-\sum_{n=1}^N x[n]^2$, is less than 6\% relative to the total energy of the original traffic $\sum_{n=1}^N x[n]^2$,
which suggests the negligible energy contributed by frequency components beyond the three main components.

\begin{figure} [bth]
\begin{center}
\includegraphics*[width=8.8cm]{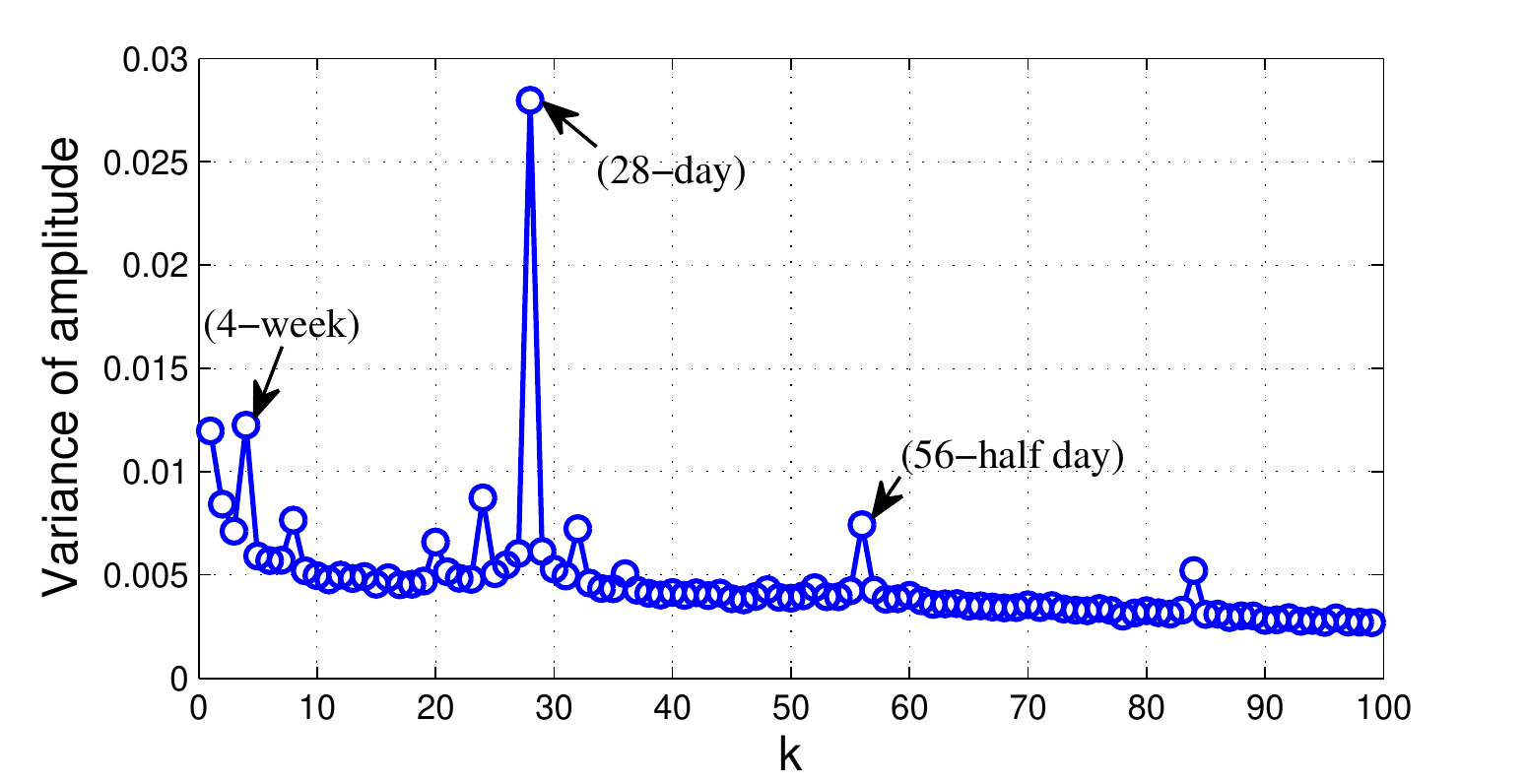}
\end{center}
\caption{Variance of the frequency components across the five identified patterns.}\label{fig:VoA}
\end{figure}

\begin{figure} [bth]
\begin{center}
\includegraphics*[width=8.8cm]{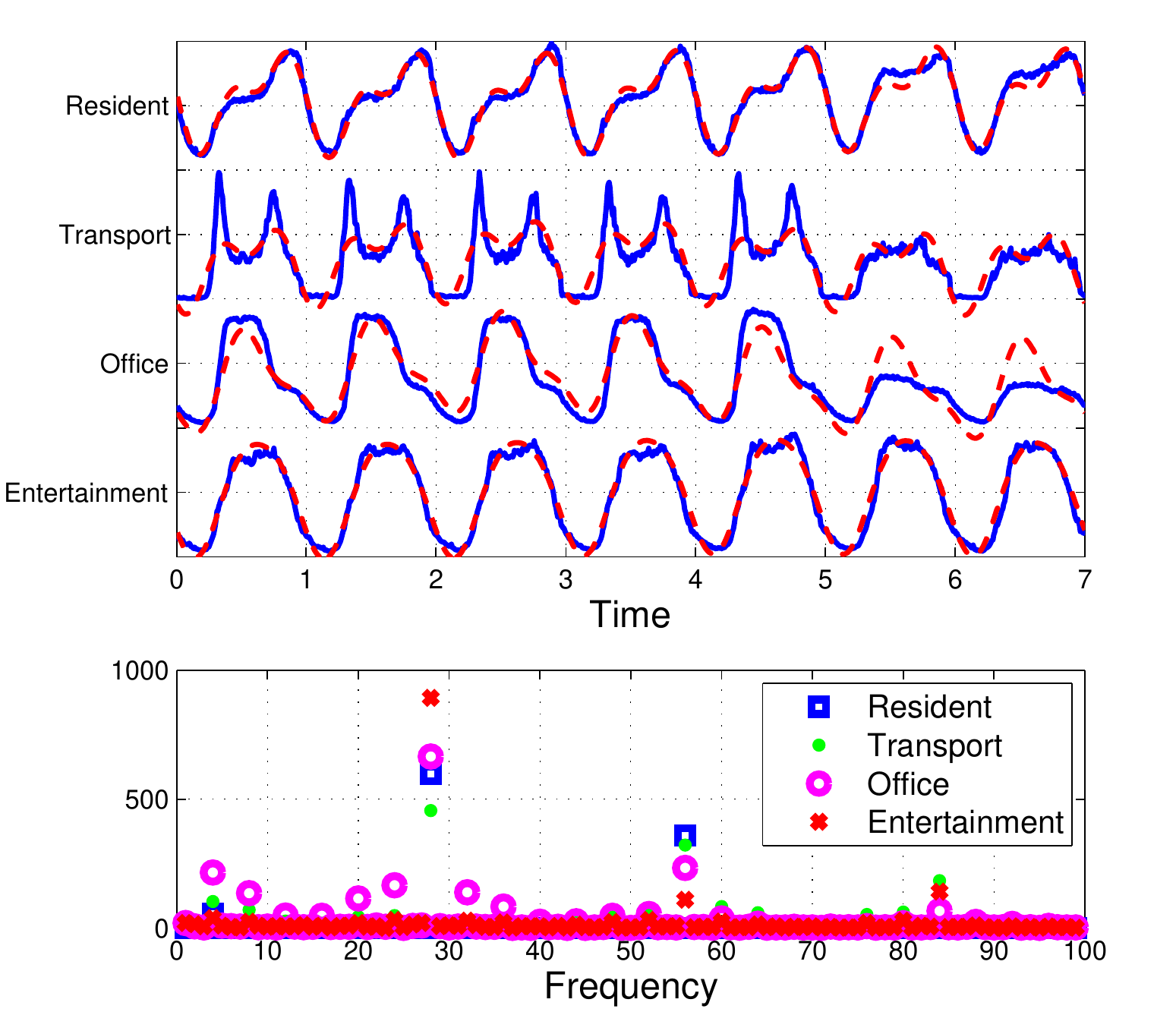}
\end{center}
\caption{Reconstructed time-domain traffic of the five patterns using the three principal frequency domain components.}\label{fig:fexa4}
\end{figure}

To further understand the capability of signal reconstruction using the three points, we analyze the variance of amplitude of DFT at each frequency component \WHD{for different cellular tower, and the result is shown in Figure~\ref{fig:VoA}.}
We can observe that the DFT variances of the three frequency components are larger compared to the rest.
In addition, we use the DFT to analyse \WHD{the aggregate traffic for cellular towers of the four primary traffic patterns} in Figure~\ref{fig:fexa4}.
We can find that the reconstructed curves are also very close to the original curves,
and their DFT spectrum varies most significantly at the three frequency components,
which suggests that these three frequencies are the most important components in
distinguishing towers of different traffic patterns
as well as constructing a time-domain traffic.

\begin{figure*}[t]
\centering
\subfigure[k=4]{\includegraphics[width=.310\textwidth]{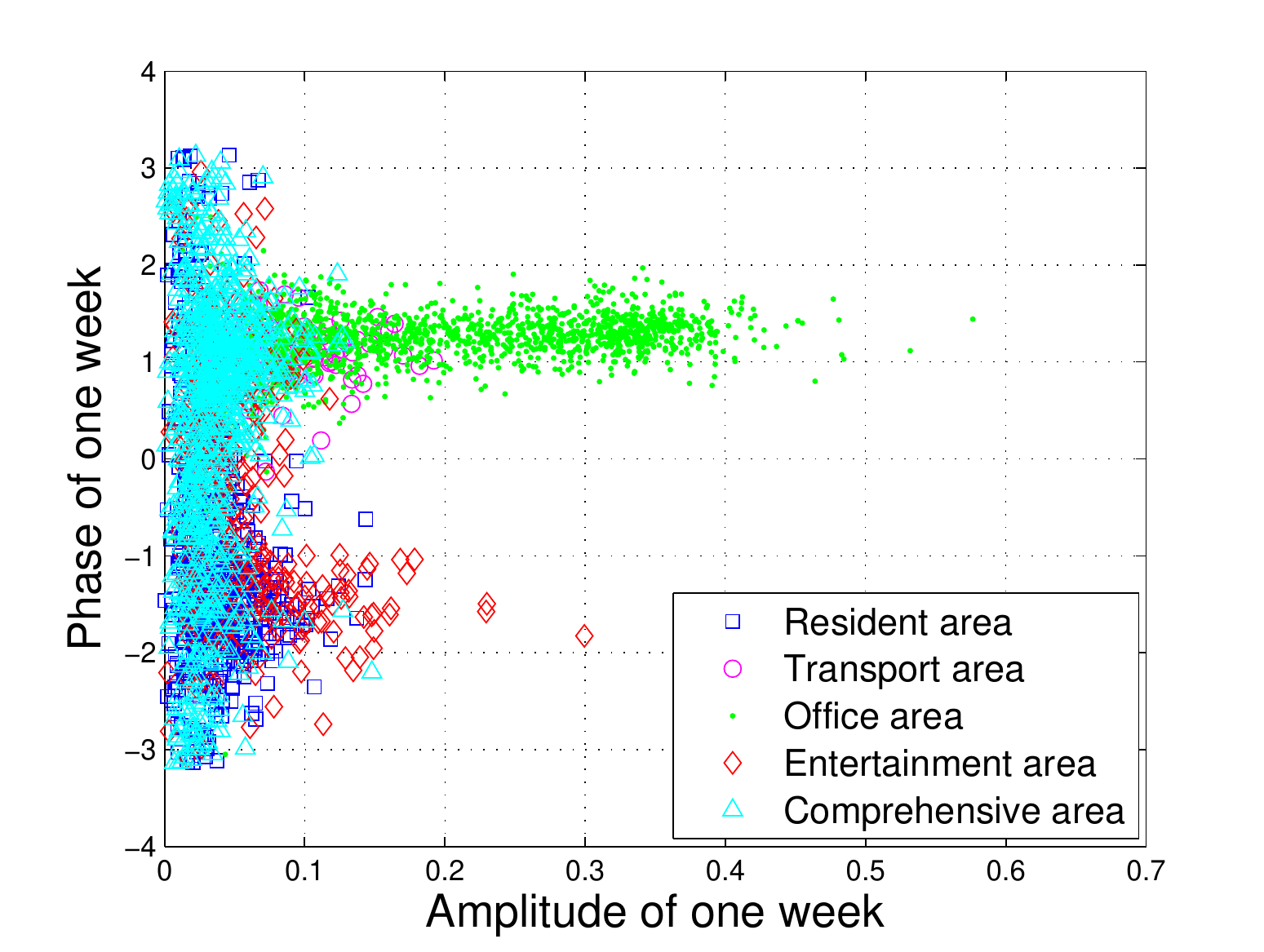}}
\subfigure[k=28]{\includegraphics[width=.310\textwidth]{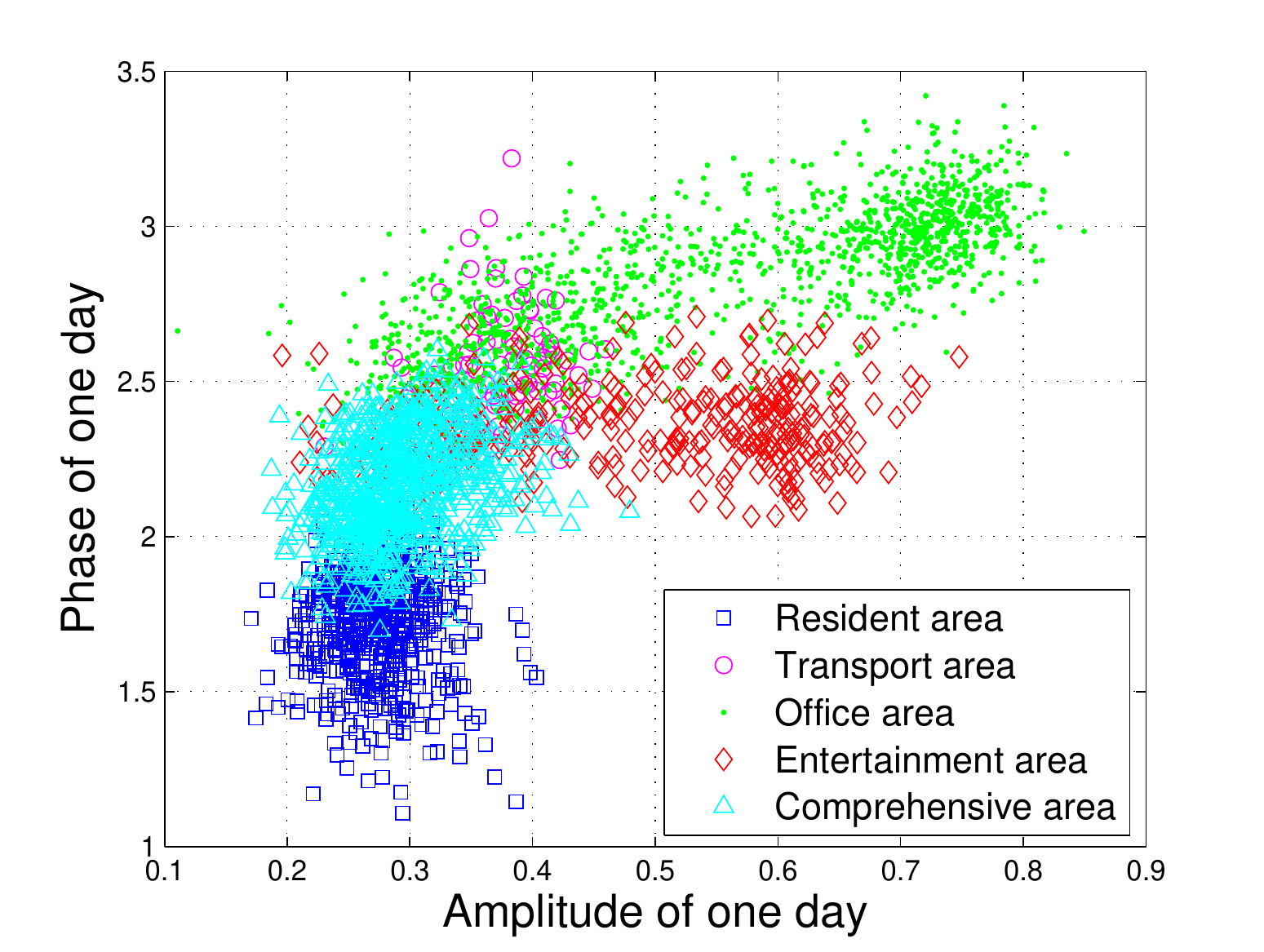}}
\subfigure[k=56]{\includegraphics[width=.310\textwidth]{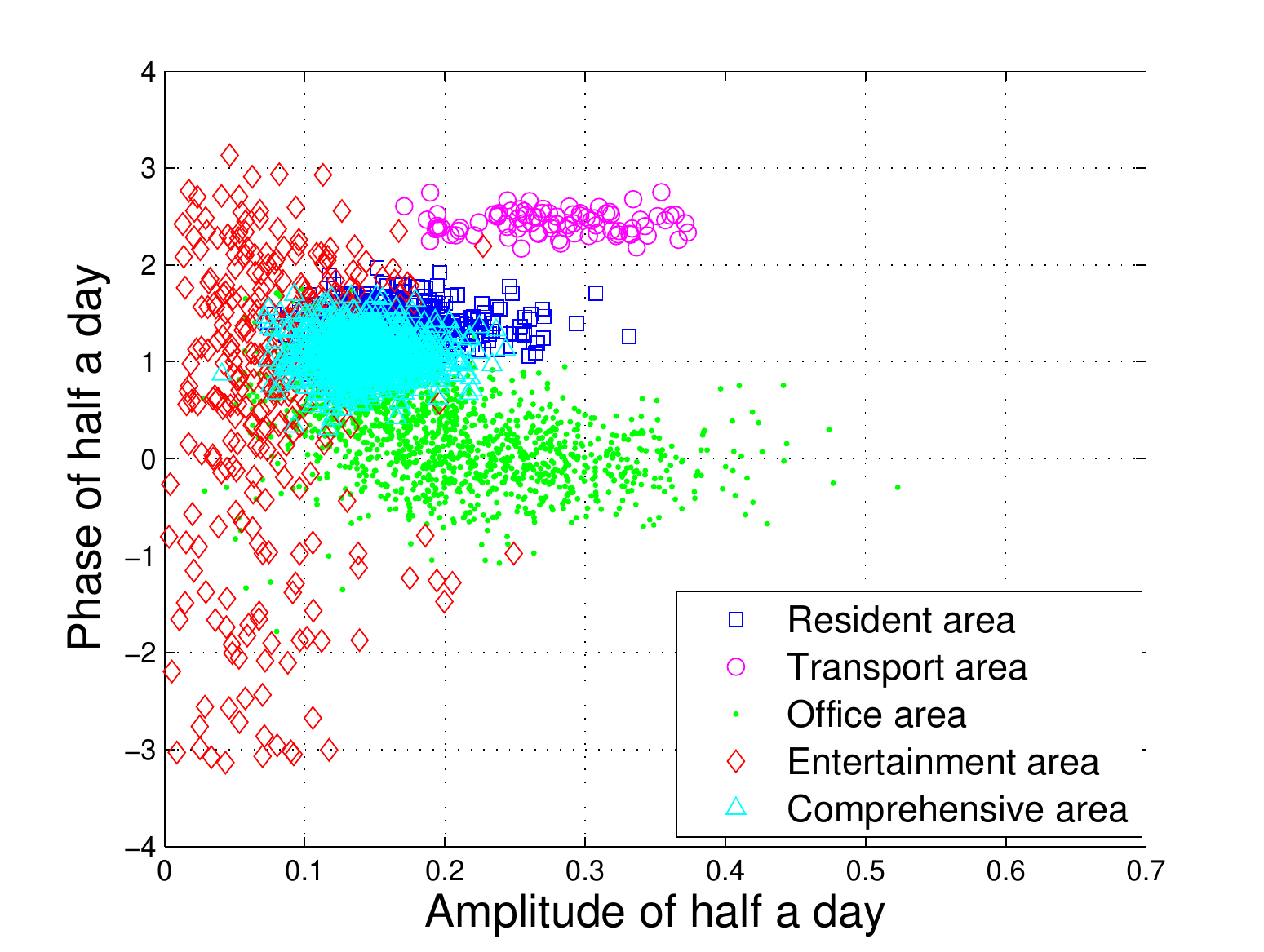}}
\caption{Phase and amplitude distribution of the three principal frequency components in the frequency domain.}\label{fig:APP}
\end{figure*}

\begin{figure*}[t]
\centering
\subfigure[Week]{\includegraphics[width=.320\textwidth]{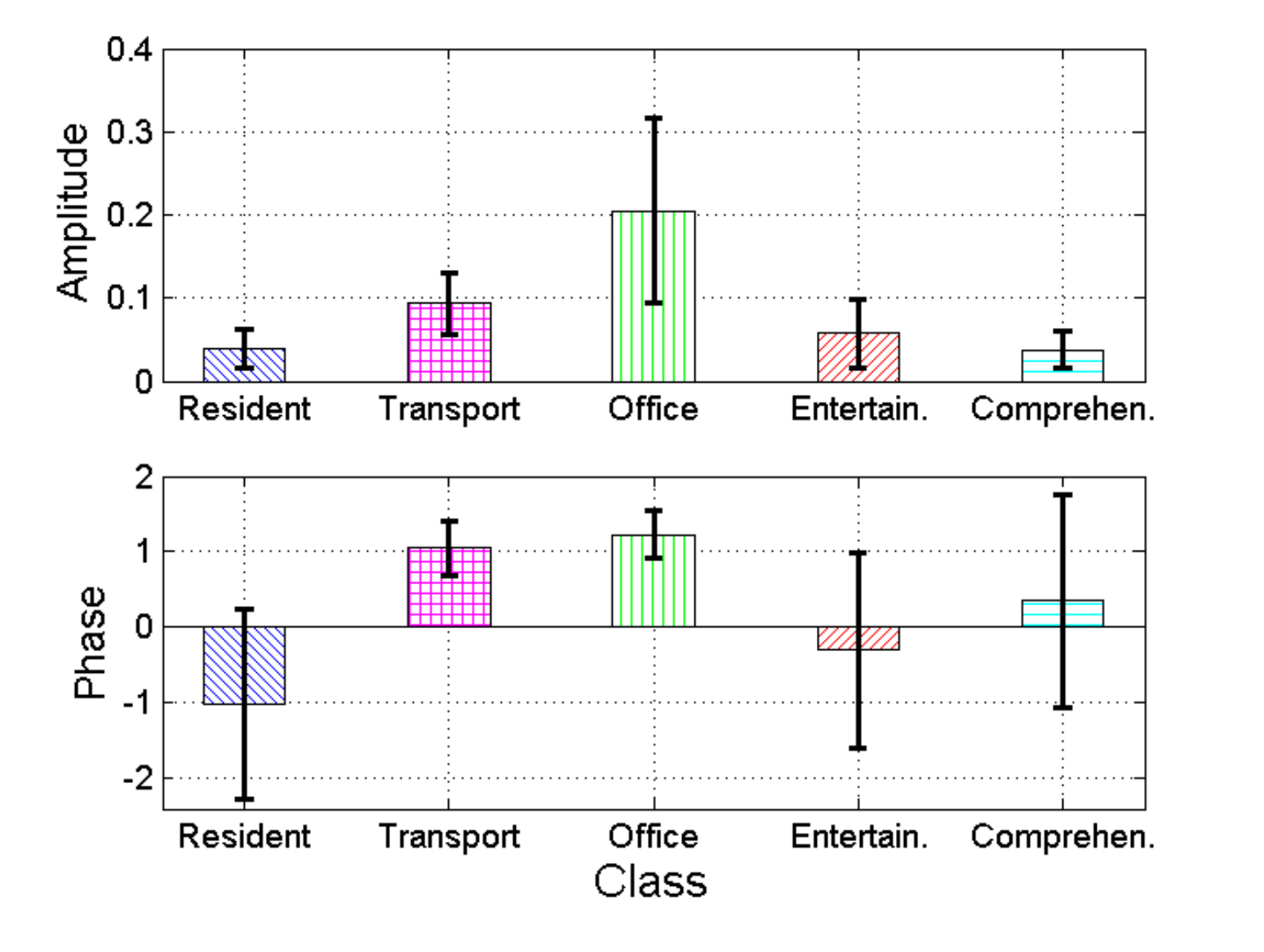}}
\subfigure[One day]{\includegraphics[width=.320\textwidth]{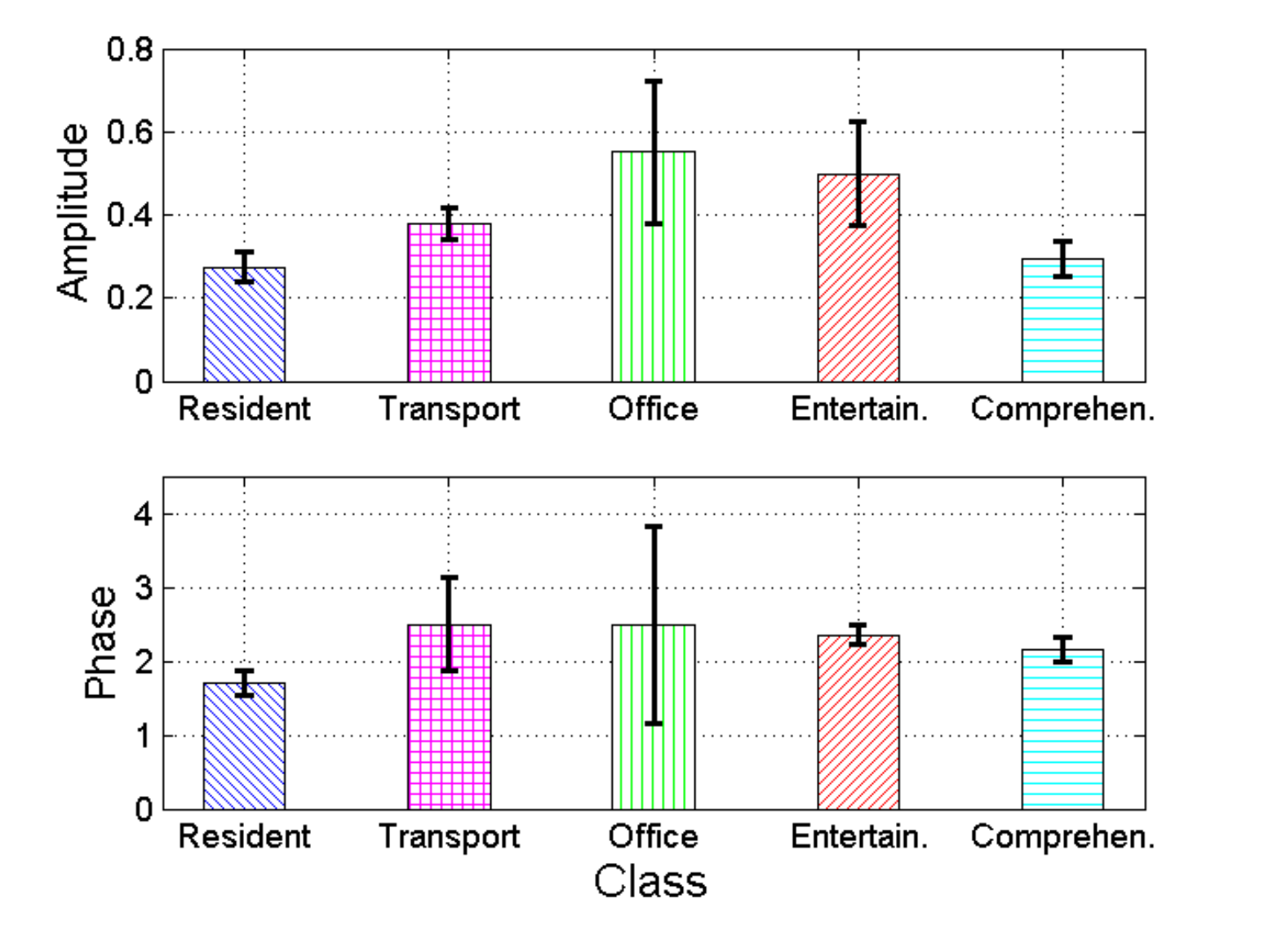}}
\subfigure[Half a day]{\includegraphics[width=.320\textwidth]{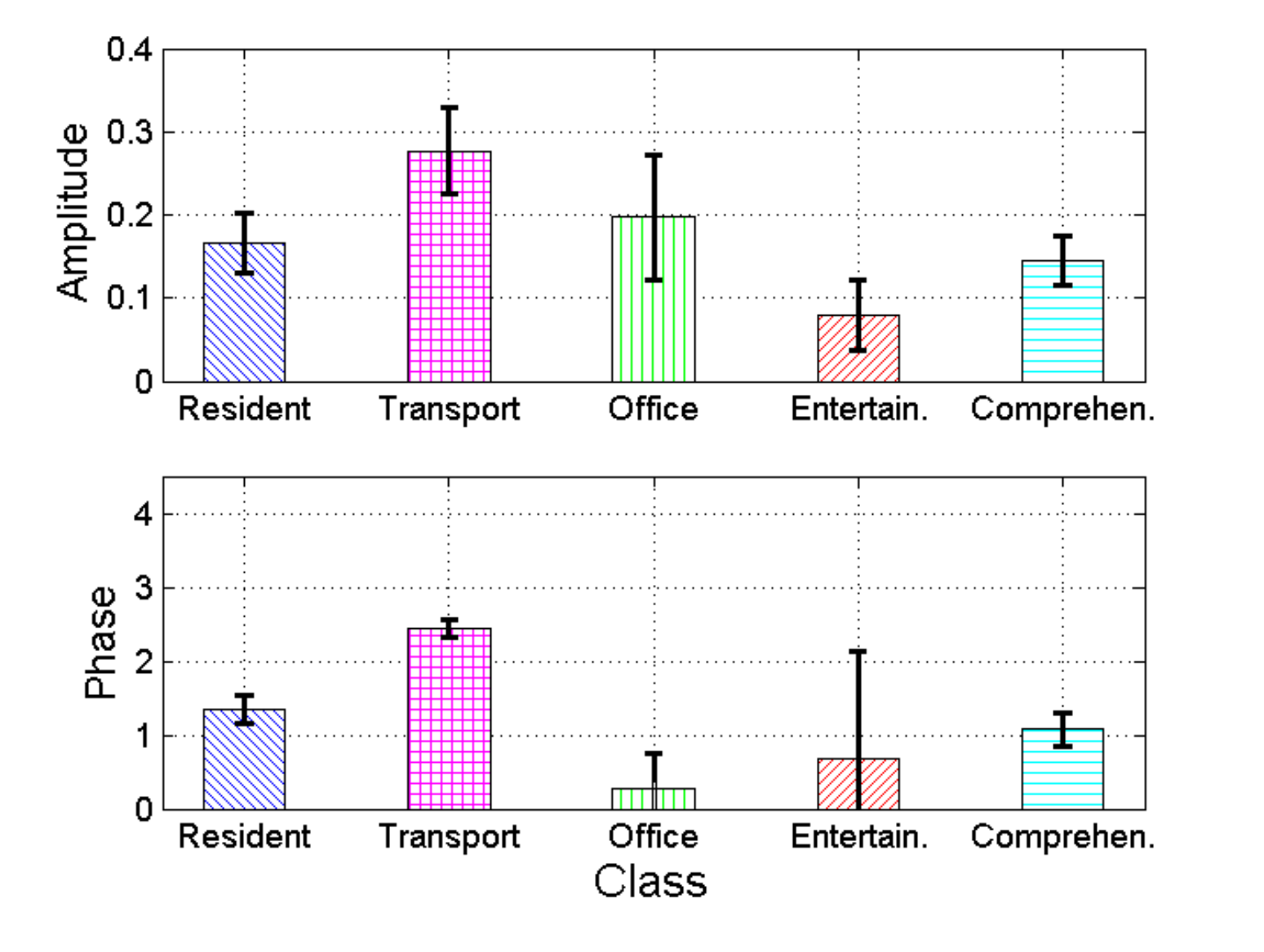}}
\caption{Means and standard deviations of amplitude and phase for cellular towers from the five identified patterns.}\label{fig:APPP}
\end{figure*}

\subsection{Visualized Analysis in Frequency Domain}
In order to better understand the five traffic patterns of towers in frequency domain,
we now provide visualized frequency analysis of them.
In addition,
based on our earlier observation in Section~5.1,
we only analyze the three frequencies corresponding to one week, one day, and half a day.
Since each DFT point is a complex number, we analyze the distribution of its amplitude and phase extracted by the following expressions:
\begin{equation*}\begin{array}{l}
\begin{cases}
A^m_k=||X^m[k]||,\\
P^m_k=arg\ X^m[k],
\end{cases}
\end{array}
\end{equation*}
where $A^m_k$ and $P^m_k$ are the amplitude and phase of DFT for tower $m$ at the $k_{th}$ frequency component.

\WHD{The larger amplitude reflects the stronger periodicity at corresponding frequency,
while different phases of DFT indicate different peak time or valley time.
Intuitively, for example, larger $A^m_{28}$ indicates the cellular tower $m$ is located at the area that is significantly influenced by the holiday at the weekend, such as office and entertainment area.
On the other hand, since the traffic peak at office area tends to be reached at weekdays,
while it at entertainment area tends to be reached at weekends,
their $P^m_{28}$ will have much difference.
Thus, by frequency analysis, we can quantify the inherent time-domain periodicity of traffic,
which is difficult to achieve by the time domain analysis.
}

Figure~\ref{fig:APP} shows the distribution of the amplitude and phase of towers deployed in the comprehensive, residential, office, transport, and entertainment areas.
Meanwhile, means and standard deviations of the amplitude and phase for towers at the three frequency components
of towers in the 4 types of areas are presented in Figure~\ref{fig:APPP}.

From Figure~\ref{fig:APP}(a) and Figure~\ref{fig:APPP}(a), we can observe that towers in office area have the strongest periodicity of one week.
Their phases mainly concentrate around 1.35,
while the phase of towers in residential and entertainment area centers around -1.65, about $\pi$ away from 1.35. This $\pi$ separation suggests that towers in residential and entertainment area have reverse traffic characteristics as that in the office area \WHD{in the scale of one week}.

As we can observe in Figure~\ref{fig:APP}(b), the distribution of towers is continuous with respect to the phase of one day.
Moreover, it shows a smooth traffic transition from residential area to comprehensive and transport area, and finally to office area.
On the other hand, according to our priori knowledge, the human migration flow usually leads to the peaks of traffic of areas appear sequentially with the same order that the flow passes through,
which coincides with our observed phenomenon.
Thus, such transition suggests the human migration flow from home to office via transport during rush hours.
In Figure~\ref{fig:APPP}(b), we can also observe that the means of their phase are incremental with the same order.

Figure~\ref{fig:APP}(c) and Figure~\ref{fig:APPP}(c) show characteristics of the amplitude and phase of the frequency component which stands for half a day.
The amplitude of this frequency component indicates the strength of double-hump characteristic.
In Figure~\ref{fig:APPP}(c), we can observe that the amplitude of towers in transport area is the largest, indicating their strongest double-hump characteristic.
This result coincides with our priori knowledge that there are two rush hours of transport area in the morning and evening, respectively.
In Figure~\ref{fig:APP}(c), we find that traffic of residential and office area are not separated by traffic of transport area. This observation is not contradictory to our pervious analysis because the directions of people commute in the morning and afternoon are reversed.

Overall, the amplitude and phase of the three frequency components show a strong capability of differentiating towers with different traffic patterns. Based on the observations, we make the following statements.
First, the most representative tower in each cluster is not the centroid. In fact, it is the farthest non-noise point from the hyperplanes, which separate clusters. To understand this problem, let us think about the points around a hyperplane, where we observe similar traffic patterns of points even though they belong to different clusters. In geographical context, these towers are deployed in areas of mixed urban functions. In contrast, the points far from the separating hyperplane are located at areas of a single urban function.
Although perhaps not the most representative points, cluster centroids can well characterize the traffic patterns since they are distant from others clusters.

Second, the frequency-domain features of towers are distributed in a polygon.
Such polygon is formed because the profile of each cluster in Figure~\ref{fig:APP} has a cigar shape.
Thus, different features of towers can be
regarded as being linear relevant or piecewise linear relevant approximately, which overlayed with a Gaussian
noise can form the cluster with the cigar shape.
As a result, a point in the frequency domain can be seen as a linear combination of the four vertex of the polygon,
$i.e.$, the four most representative points, which we call as the four primary components.

\begin{figure} [t]
\begin{center}
\includegraphics*[width=7.5cm]{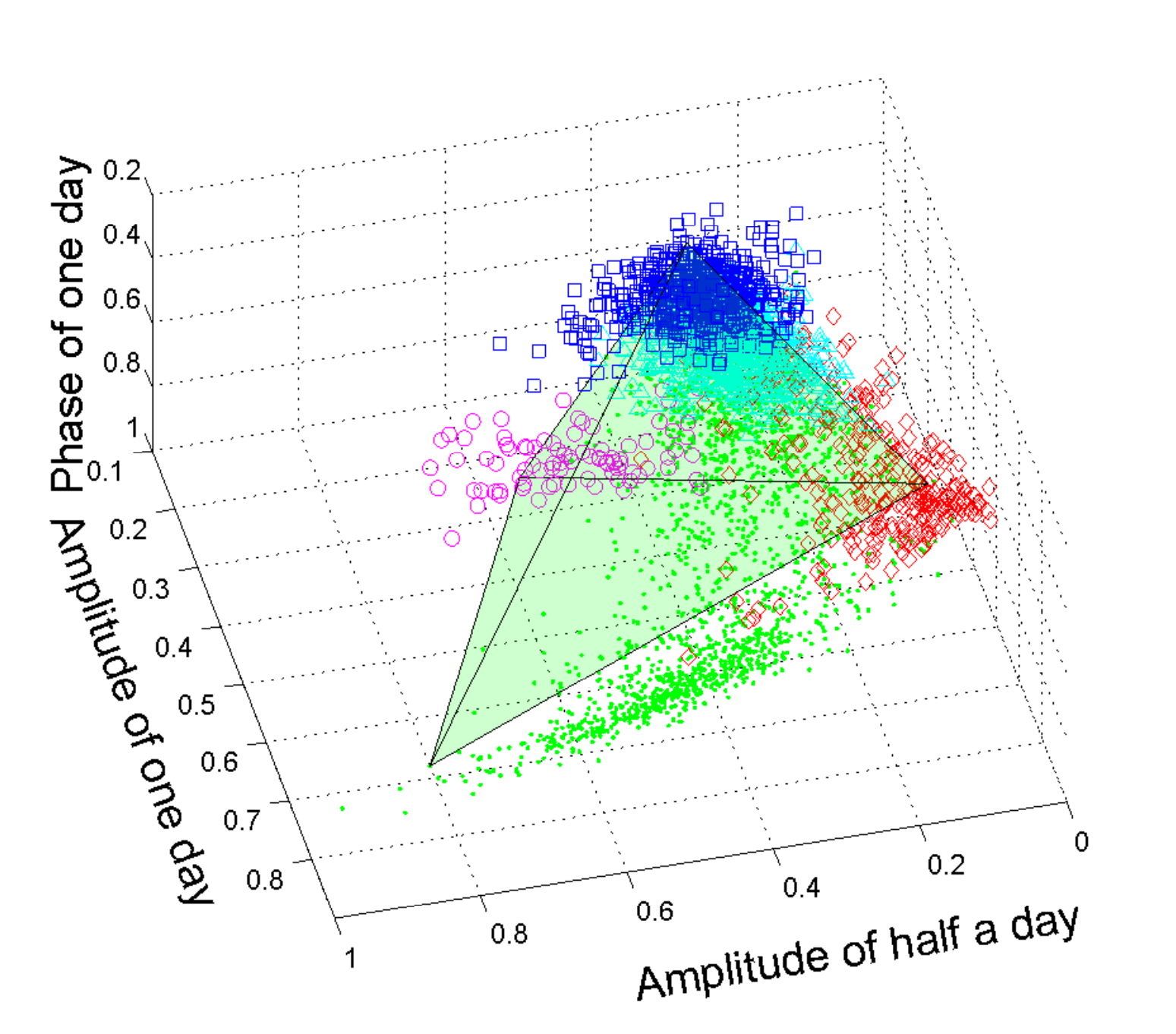}
\end{center}
\caption{Three-dimensional view of the distribution of cellular towers in the frequency domain.}\label{fig:pol}
\end{figure}

To illustrate these two statements, we plot the distribution of towers and corresponding polygon in Figure~\ref{fig:pol}. For better understanding, we only show three features, including amplitude and phase of one day, and amplitude of half day. According to our first statement, the most representative tower in each cluster is the furthest one from the hyperplane.
\WHD{Specifically, we do not calculate the hyperplanes, and only search for points with largest distance from points of other clusters.}
In addition, we use the density of the towers, i.e., the number of towers within a fixed distance away from \WHD{it} in the feature space, as a decision function to ensure that \WHD{the tower} is not a noise point.
Figure~\ref{fig:pol} shows that all the towers are distributed in or along the edge and plane of the polygon, as we discussed above.

\subsection{Component Analysis of Cellular Towers in Comprehensive Area}

\WHD{Based on the statements above,
we may use a linear combination of the four most representative cellular towers to present each point in the polygon.} By looking at the coefficient of each primary component, we can obtain the percentage of corresponding urban function of the area where an arbitrary cellular tower is deployed. We formulate the process of obtaining the coefficients as a quadratic programming problem, which is shown below:
\begin{equation*}
\setlength{\abovedisplayskip}{0pt}
\setlength{\belowdisplayskip}{0pt}
\begin{array}{l}
\text{minimize}\ \ \ ||F-F^r||^2 \\
\text{subject to}\ \
\begin{cases}
&\!\!\!\! \sum_{i=1}^4 F^0_i x_i =F^r, \\
&\!\!\!\! \sum_{i=1}^4 x_i =1,  \\
&\!\!\!\! x_i\geq 0, \ \ \ i= {1,...,4},
\end{cases}
\end{array}
\end{equation*}
where ||{\boldmath$\cdot$}|| is the 2-norm of a vector, $F$ is the feature of the target tower, $F^0_i$ is the feature of the most representative tower for cluster $i$ in the frequency domain, and $x_i$ is the obtained coefficient for cluster $i$. In this example, the feature of tower $m$, $F^m$, is $(A_{28}^m,P_{28}^m,A_{56}^m)$,
where $A_{28}^m$ ,$P_{28}^m$, $A_{56}^m$ are the amplitude of one day, phase of one day, and amplitude of half a day for tower $m$, respectively.
We use the quadratic programming to solve the problem because the traffic of an actual tower is usually overlayed with various noises, such that these points close to the plane of the the polygon may be driven out of the polygon.
By solving this quadratic programming,
for points inside the polygon, we can find their exact convex combinations,
while for some point outside the polygon, we can find the point in the polygon with the smallest distance to the target point, which is a good approximation.

We dedicatedly select a list of towers in the comprehensive area.
Then, we use the method presented above to solve the convex combinatorial coefficients of them.
We compare these coefficients with a transform of the previously introduced POI, 
$i.e.$, the term frequency-inverse document frequency (TF-IDF) of the corresponding types and locations.
TF-IDF is a numerical statistic that is intended to reflect how important a word is to a document.
\WHD{Similarly, it is used to reflect how important the POI of a specific type is in our analysis,
which has been proposed in existing works,
i.e., Yuan $et\ al.$ \cite{yuan2012discovering} provided a TF-IDF-based method to cluster regions of different functions, which solely uses the POI data.}
Specifically, TF-IDF can be calculated as the following:
\begin{equation*}\begin{array}{l}
\setlength{\abovedisplayskip}{0pt}
\setlength{\belowdisplayskip}{0pt}
\begin{cases}
\setlength{\abovedisplayskip}{0pt}
\setlength{\belowdisplayskip}{0pt}
\text{IDF}_i=\text{log}(M/M_i),\\
\text{TF-IDF}^m_i=\text{IDF}_i\cdot\text{log}(1+\text{POI}^m_i),
\end{cases}
\end{array}
\end{equation*}
where $M$ is the total number of towers, and $M_i$ is the number of towers of which the POI of type $i$ appears within a specific distance, 
and POI$^m_i$ is the times that the POI of type $i$ appears within a fixed distance of tower $m$.
To be better compared with, we normalize the TF-IDF of each tower by the sum of TF-IDF of all the four types for this tower, which is called as the normalized TF-IDF (NTF-IDF). This process can be formulated as the following:
\begin{equation*}
\setlength{\abovedisplayskip}{0pt}
\setlength{\belowdisplayskip}{0pt}
\text{NTF-IDF}^m_i=\text{TF-IDF}^m_i/\sum_{j=1}^4\text{TF-IDF}^m_j.
\end{equation*}
The obtained NTF-IDF is proportional to the POI for each type,
which roughly represents the density of the corresponding function in the corresponding area.
\WHD{Specifically, NTF-IDF close to 0 indicates this area do not have the corresponding function.
However, the largest NTF-IDF do not completely indicate the corresponding function is dominant in the area,
since it is also influenced by the size of related points and corresponding distance.
For example, a large and close subway station has more influence than a small and far residential building on a cellular tower.}

Then, the result is shown in Table~\ref{tab:CvsPOI}. \WHD{Expect for the towers in the comprehensive area, the NTF-IDF of the four most representative towers is also provided in the table.
We can observe that their NTF-IDF of corresponding types is much larger than others,
which is very close to 1, indicating the areas where they are located have a single type of function.
As for towers in the comprehensive area,
There are multiple relative large NTF-IDF for a cellular tower.
As discussed earlier, this may lead to inaccuracy since the influence of the size of related points and corresponding distance.
Thus, we only consider the consistency of the small NTF-IDF and combination coefficients.
We can observe that} the majority of the smallest NTF-IDF$^m_i$ in all $m$ for some fix $i$ corresponds to the smallest coefficient in all $m$ for the same $i$, respectively.
For example, \WHD{NTF-IDF$^{P3}_2$} and \WHD{NTF-IDF$^{P4}_1$} are 0, and their corresponding coefficients are also 0.
Thus, the obtained convex combination coefficients coincide with the POI distribution,
indicating the correctness of our theory.

\begin{table}[t]
\centering
\caption{\WHD{Convex combination coefficients and NTF-IDF.}}\label{tab:CvsPOI}
\begin{center}
\begin{tabular}{|p{0.6cm}<{\centering}|p{0.5cm}<{\centering}p{0.5cm}<{\centering}p{0.5cm}<{\centering}p{0.5cm}<{\centering}|p{0.5cm}<{\centering}p{0.5cm}<{\centering}p{0.5cm}<{\centering}p{0.5cm}<{\centering}|}\hline
\multirow{2}{*}{} & \multicolumn{4}{c|}{Coefficient} & \multicolumn{4}{c|}{NTF-IDF} \\ \cline{2-9}
& \multicolumn{1}{c}{1} & \multicolumn{1}{c}{2} & \multicolumn{1}{c}{3} & \multicolumn{1}{c|}{4} & \multicolumn{1}{c}{1} & \multicolumn{1}{c}{2} & \multicolumn{1}{c}{3} & \multicolumn{1}{c|}{4}\\ \hline
\WHD{F1}     &\cellcolor[rgb]{.9,.7,.7}1.00 &0.00 &0.00 &0.00     &\cellcolor[rgb]{.9,.7,.7}1 &0 &0 &0\\
\WHD{F2}     &0.00 &\cellcolor[rgb]{.9,.7,.7}1.00 &0.00 &0.00     &0.00 &\cellcolor[rgb]{.9,.7,.7}0.81 &0.05 &0.14\\
\WHD{F3}     &0.00 &0.00 &\cellcolor[rgb]{.9,.7,.7}1.00 &0.00     &0.00 &0.00 &\cellcolor[rgb]{.9,.7,.7}1.00 &0.00\\
\WHD{F4}      &0.00 &0.00 &0.00 &\cellcolor[rgb]{.9,.7,.7}1.00    &0.00 &0.00 &0.28 &\cellcolor[rgb]{.9,.7,.7}0.72\\
\hline\hline
P1     &\cellcolor[rgb]{.9,.7,.7}0.79 &0.13 &0.08 &\cellcolor[rgb]{.7,.9,.9}0.00     &0.44 &0.36 &0.04 &\cellcolor[rgb]{.7,.9,.9}0.16\\
P2     &0.09 &0.07   &\cellcolor[rgb]{.7,.9,.9}0.00    &\cellcolor[rgb]{.9,.7,.7}0.84   &0.36    &0.39    &\cellcolor[rgb]{.7,.9,.9}0.03    &0.22\\
P3 &0.23 &\cellcolor[rgb]{.7,.9,.9}0.00 &0.15    &0.62  &\cellcolor[rgb]{.9,.7,.7}0.68         &\cellcolor[rgb]{.7,.9,.9}0.00    &0.07    &0.25\\
P4  &\cellcolor[rgb]{.7,.9,.9}0.00    &\cellcolor[rgb]{.9,.7,.7}0.29   &\cellcolor[rgb]{.9,.7,.7}0.42    &0.29  &\cellcolor[rgb]{.7,.9,.9}0.00   &\cellcolor[rgb]{.9,.7,.7}0.68    &\cellcolor[rgb]{.9,.7,.7}0.07    &0.24\\
P5     &0.35 &0.18 &0.22 &0.25     &0.12 &0.55 &0.05 &\cellcolor[rgb]{.9,.7,.7}0.28\\
\hline
\end{tabular}
\end{center}
\end{table}

\begin{figure} [t!]
\begin{center}
\includegraphics*[width=7.5cm]{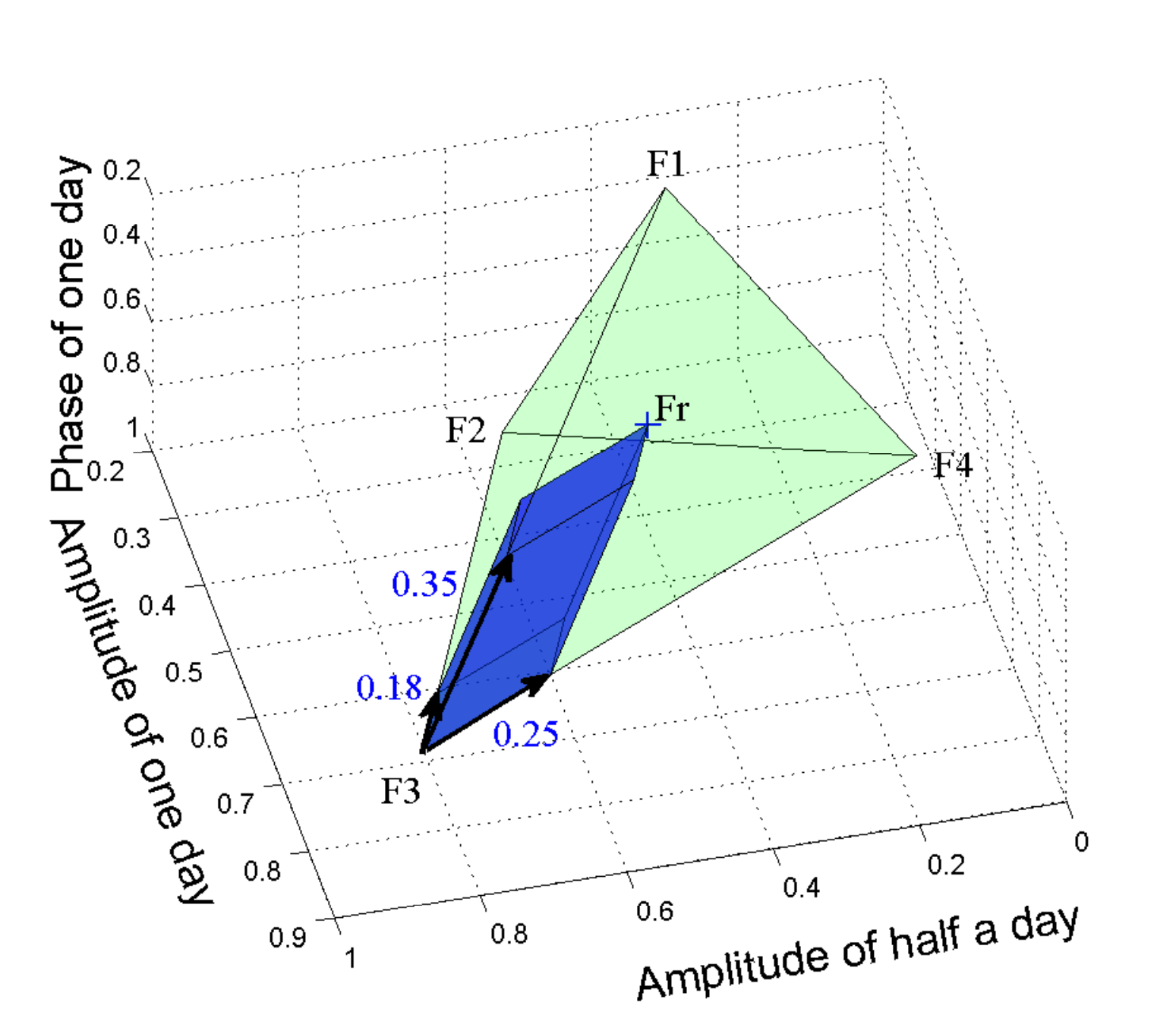}
\end{center}
\caption{Convex combination for P5 \WHD{in Table~\ref{tab:CvsPOI}} in frequency domain.}\label{fig:cmb}
\end{figure}

\begin{figure} [t!]
\begin{center}
\includegraphics*[width=8.8cm]{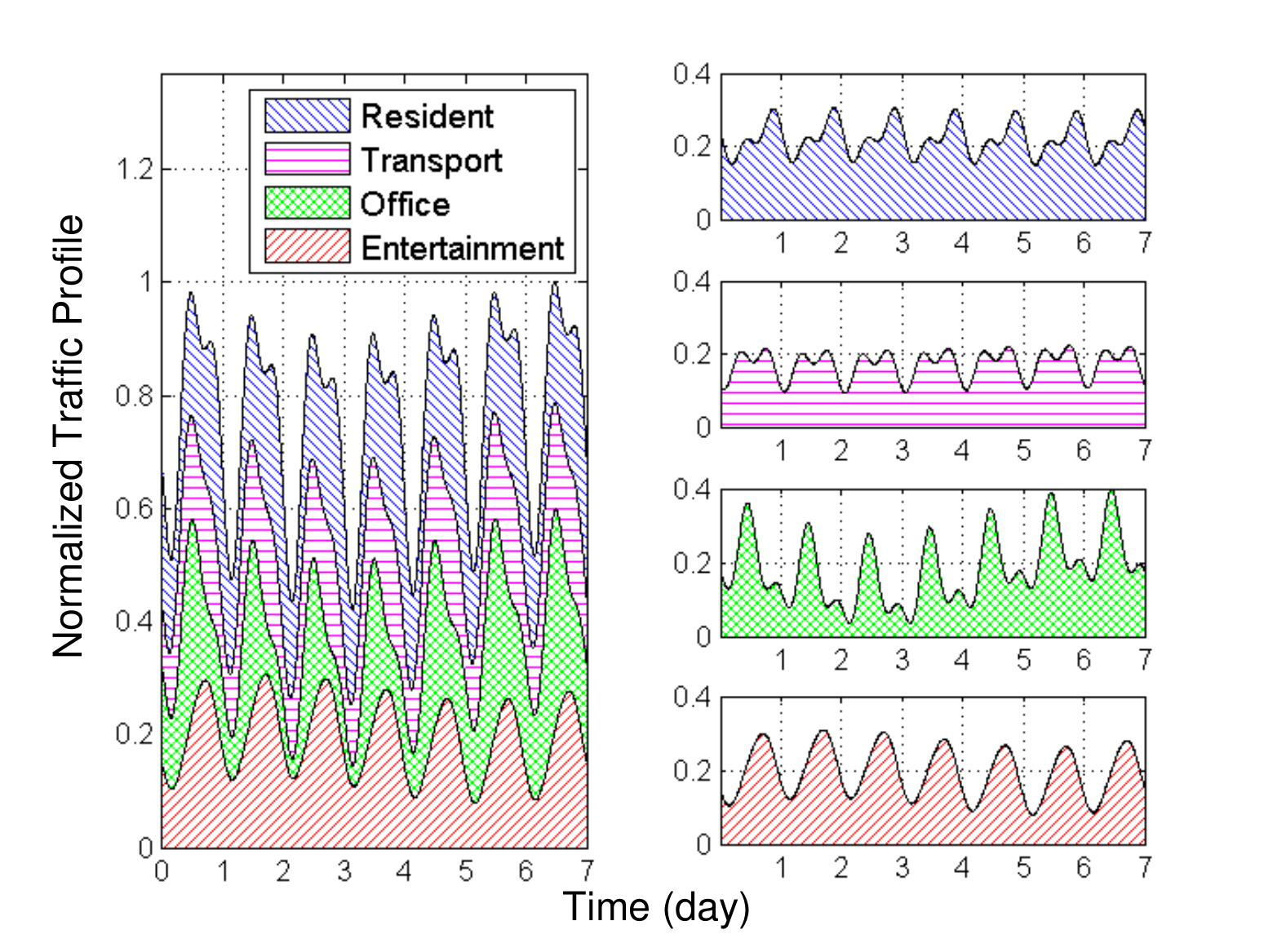}
\end{center}
\caption{Convex combination for P5 in time domain.}\label{fig:pcmbt}
\end{figure}

To further illustrate the convex combination, we take the tower P5 in Table~\ref{tab:CvsPOI} as an example, and show its combination of frequency and time domain, respectively in Figure~\ref{fig:cmb} and Figure~\ref{fig:pcmbt}.
For a point inside the polygon, we can find its exact convex combination, that is:
\begin{equation*}
\setlength{\abovedisplayskip}{0pt}
\setlength{\belowdisplayskip}{0pt}
F=F^r=\sum_{i=1}^4F^0_i x_i=F^0_3+\sum_{i=1,i\not= 3}^4x_i(F^0_i-F^0_3).
\end{equation*}
As shown in Figure~\ref{fig:cmb},
in the feature space, the vector $(0,Fr)$ can be divided to the vector $(0,F3)$
and the weighted sum of the vector $(F3,F1),(F3,F2),(F3,F4)$.
For P5, the weights are 0.35, 0.18 and 0.25, respectively, which is just as coefficients of cluster 1, 2, 4 in Table~\ref{tab:CvsPOI} of P5.

On the other hand, in Figure~\ref{fig:pcmbt}, we show the components of traffic corresponding to four primary clusters for the comprehensive tower P5.
Areas of different colors in the left figure represent components of different primary traffic patterns. To be better distinguished, each component is added with a static bias.
In addition, we plot each component individually in the right figure.
The result indicates that traffic patterns of an arbitrarily cellular tower can be approximated by a convex combination of four primary traffic patterns. The size of each component is highly related to density of corresponding function around the tower.
It further demonstrates the correctness and usefulness of our frequency analysis method.

\section{Related Work}

The digital footprints of human activities and network behaviors contributed by mobile devices have led to a plethora of investigations on the intersection between human and network dynamics\cite{kosinski2013private,laurila2012mobile}.  This section summarizes relevant research from three perspectives --- data sources, types of collected data and targeted applications.

Dataset collected from mobile devices for investigating human behaviors and network performance can be divided into two broad categories: (1) data collected from mobile devices and (2) traces collected by mobile operators \cite{cici2015decomposition}. For the first categories, users or experimenters report their semantically annotated data about the locations, phone usages and network performance by installing some Apps in their devices\cite{noulas2013exploiting, hu2014quality}. The limitation of this approach comes from the limited number of users sampled, which cannot stand for the global characteristics of a large scale cellular network. On the other hand, in the dataset collected by cellular operators, users are passively monitored and the operators decide which information to collect\cite{gonzalez2008understanding,toole2012inferring}. As a result, the collected data is continuous as long as devices are connected, and includes detailed information of users behaviors, such as duration of each Internet connection. As a result, data collected via the second approach enables the study of overall network behaviors, such as large scale of human mobility and call activities analysis. In this paper, we use the data collected by an ISP for investigating the traffic patterns of large scale cellular towers.

Extensive studies have used various types of cellular data for understanding the characteristics of large scale cellular towers. For example, cell phone activities, commonly know as Call Description Records (CDR), are used for capturing human communication activities\cite{cici2013quantifying}. In addition, it is also used for recovering the human mobility trajectory\cite{gonzalez2008understanding}, inferring demographics\cite{dong2014inferring}, and uncovering urban ecology\cite{cici2015decomposition}. Another type of data is the device-level metric obtained from mobile devices, such as device and application usage\cite{falaki2010first, shafiq2012characterizing}, network access bandwidth\cite{hu2014quality}, energy computation\cite{hu2015energy}, personal GPS locations\cite{zheng2008understanding}, etc. With the popularity of 3G and LTE access, mobile and application data traces become available as well. Cici et al. \cite{cici2015decomposition} characterizes the relationship between people's application interests and mobility patterns based on a population of over 280, 000 users of a 3G mobile network. Lee et al. \cite{lee2014spatial} demonstrated that the spatial distribution of the traffic density can be approximated by the log-normal or Weibull distribution. However, mobile data traffic across a city-wide range with different time scale and variations  contains complicated interaction between the space and time, which requires a  deep and comprehensive understanding. The analysis and models in this work provide such insights.

Cellular network traces have been used for enabling a set of applications. The footprint of mobile devices and cellular network has been used to model human mobility and trajectories \cite{gonzalez2008understanding, eagle2009inferring, toole2012inferring, das2014contextual}. Barabasi et al. \cite{gonzalez2008understanding,song2010limits,song2010modelling} studied the mobility behaviors of 100k mobile users by analyzing the CDR data, and found that the trajectories of human is not as random as previously proposed levy flight or random walk models. Instead, it presents a high confidence of predictability\cite{song2010limits}, and temporal and spatial regularity\cite{gonzalez2008understanding}.
The cellular network traces have also been used for characterizing and modelling the cellular data traffic patterns. Shafiq et al. \cite{Zubair2011Characterizing} modelled the internet traffic dynamics of cellular devices. Jin et al. \cite{jin2012characterizing} characterized data usage patterns in large cellular network. And Zhang et al. \cite{Zhang2012Understanding} tried to understand the characteristics of cellular data traffic by comparing it to wireline data traffic.
Other studies combine the CDR, GPS locations, and application traces to investigate the land usage\cite{toole2012inferring,pei2014new}, social interactions\cite{eagle2009inferring}, location-based patterns\cite{das2014contextual}, and web and data access patterns\cite{keralapura2010profiling,jin2012characterizing}.
In this paper, we focus on investigating the mobile data traffic patterns from different domains, including time, location and frequency, which provides a comprehensive understanding of the traffic patterns of large scale cellular towers with a simple but deep model that is able to characterize the city \WHD{geographical} features and human communication regularity.

In conclusion, we study a large scale urban mobile data access traces collected by the commercial mobile operators involving over 9600 towers and 150,000 subscribers. We first design an analysis framework for processing large scale cellular traffic data. Then, we reveals the basic but fundamental patterns embedded in thousands of cellular towers, which paves a way toward a comprehensive understanding of the connection among mobile data traffic, urban ecology and human behaviors.

\section{Conclusions}
In this paper, we carry out, to the best of our knowledge, the first study of traffic patterns embedded in large scale 3G and LTE towers deployed in the urban environment. We propose a powerful model which combines time, location and frequency information for analyzing the traffic patterns of thousands of cellular towers. Our analysis reveals that the dynamic urban mobile traffic usage exhibits only five basic time domain patterns. In addition, the traffic of any tower can be reconstructed accurately using a linear combination of four primary components corresponding to human activity behaviors. Our analysis provides a systematic and comprehensive understanding of dynamic and complicated mobile traffic, and opens a set of new research directions.

\section*{Acknowledgment}
This work is supported by National Basic Research Program of China (973 Program) (No. 2013CB329105) and National Nature Science Foundation of China (No. 61301080, No. 91338203, No. 91338102, and No. 61321061).

\footnotesize
\bibliographystyle{abbrv}

\begin{thebibliography}{10}

\vspace{0.2cm}
\bibitem{cisco_cellular_forcast}
Cisco visual networking index: Global mobile data traffic forecast.
\newblock 2014.

\bibitem{cici2015decomposition}
B.~Cici, M.~Gjoka, A.~Markopoulou, and C.~T. Butts.
\newblock On the decomposition of cell phone activity patterns and their
  connection with urban ecology.
\newblock In {\em Proc. of ACM MobiHoc}, pages 317--326, 2015.

\bibitem{cici2013quantifying}
B.~Cici, A.~Markopoulou, E.~Fr{\'\i}as-Mart{\'\i}nez, and N.~Laoutaris.
\newblock Quantifying the potential of ride-sharing using call description
  records.
\newblock In {\em Proc. of ACM HotMobile}, page~17, 2013.

\bibitem{corpet1988multiple}
F.~Corpet.
\newblock Multiple sequence alignment with hierarchical clustering.
\newblock {\em Nucleic acids research}, 16(22):10881--10890, 1988.

\bibitem{das2014contextual}
A.~K. Das, P.~H. Pathak, C.-N. Chuah, and P.~Mohapatra.
\newblock Contextual localization through network traffic analysis.
\newblock In {\em Proc. of IEEE INFOCOM}, pages 925--933, 2014.

\bibitem{dong2014inferring}
Y.~Dong, Y.~Yang, J.~Tang, Y.~Yang, and N.~V. Chawla.
\newblock Inferring user demographics and social strategies in mobile social
  networks.
\newblock In {\em Proc. of ACM SIGKDD}, pages 15--24, 2014.

\bibitem{eagle2009inferring}
N.~Eagle, A.~S. Pentland, and D.~Lazer.
\newblock Inferring friendship network structure by using mobile phone data.
\newblock {\em Proc. of the National Academy of Sciences},
  106(36):15274--15278, 2009.

\bibitem{falaki2010first}
H.~Falaki, D.~Lymberopoulos, R.~Mahajan, S.~Kandula, and D.~Estrin.
\newblock A first look at traffic on smartphones.
\newblock In {\em Proc. of ACM IMC}, pages 281--287, 2010.

\bibitem{gonzalez2008understanding}
M.~C. Gonzalez, C.~A. Hidalgo, and A.-L. Barabasi.
\newblock Understanding individual human mobility patterns.
\newblock {\em Nature}, 453(7196):779--782, 2008.

\bibitem{hu2014quality}
W.~Hu and G.~Cao.
\newblock Quality-aware traffic offloading in wireless networks.
\newblock In {\em Proc. of ACM MobiHoc}, pages 277--286, 2014.

\bibitem{hu2015energy}
W.~Hu and G.~Cao.
\newblock Energy-aware video streaming on smartphones.
\newblock In {\em Proc. of IEEE INFOCOM}, 2015.


\bibitem{jin2012characterizing}
Y.~Jin, N.~Duffield, A.~Gerber, P.~Haffner, W.-L. Hsu, G.~Jacobson, S.~Sen,
  S.~Venkataraman, and Z.-L. Zhang.
\newblock Characterizing data usage patterns in a large cellular network.
\newblock In {\em ACM CellNet Workshop}, pages 7--12, 2012.

\bibitem{keralapura2010profiling}
R.~Keralapura, A.~Nucci, Z.-L. Zhang, and L.~Gao.
\newblock Profiling users in a 3g network using hourglass co-clustering.
\newblock In {\em Proc. of ACM MobiCom}, pages 341--352, 2010.

\bibitem{kosinski2013private}
M.~Kosinski, D.~Stillwell, and T.~Graepel.
\newblock Private traits and attributes are predictable from digital records of
  human behavior.
\newblock {\em Proc. of the National Academy of Sciences},
  110(15):5802--5805, 2013.

\bibitem{laurila2012mobile}
J.~K. Laurila, D.~Gatica-Perez, I.~Aad, O.~Bornet, T.-M.-T. Do, O.~Dousse,
  J.~Eberle, M.~Miettinen, et~al.
\newblock The mobile data challenge: Big data for mobile computing research.
\newblock In {\em Pervasive Computing}, number EPFL-CONF-192489, 2012.

\bibitem{lee2014spatial}
D.~Lee, S.~Zhou, X.~Zhong, Z.~Niu, X.~Zhou, and H.~Zhang.
\newblock Spatial modeling of the traffic density in cellular networks.
\newblock {\em Wireless Communications, IEEE}, 21(1):80--88, 2014.

\bibitem{Zubair2011Characterizing}
J.~A. X. L. J.~W. M.~Zubair, Shafiq~Lusheng.
\newblock Characterizing and modeling internet traffic dynamics of cellular
  devices.
\newblock {\em Performance evaluation review}, 39(1):305--316, 2011.




\bibitem{maulik2002performance}
U.~Maulik and S.~Bandyopadhyay.
\newblock Performance evaluation of some clustering algorithms and validity
  indices.
\newblock {\em Pattern Analysis and Machine Intelligence, IEEE Transactions
  on}, 24(12):1650--1654, 2002.


\vfill\eject

\bibitem{noulas2013exploiting}
A.~Noulas and C.~Mascolo.
\newblock Exploiting foursquare and cellular data to infer user activity in
  urban environments.
\newblock In {\em Proc. of IEEE MDM}, volume~1, pages 167--176, 2013.

\bibitem{pei2014new}
T.~Pei, S.~Sobolevsky, C.~Ratti, S.-L. Shaw, T.~Li, and C.~Zhou.
\newblock A new insight into land use classification based on aggregated mobile
  phone data.
\newblock {\em International Journal of Geographical Information Science},
  28(9):1988--2007, 2014.

\bibitem{shafiq2012characterizing}
M.~Z. Shafiq, L.~Ji, A.~X. Liu, J.~Pang, and J.~Wang.
\newblock Characterizing geospatial dynamics of application usage in a 3g
  cellular data network.
\newblock In {\em Proc. of IEEE INFOCOM}, pages 1341--1349, 2012.

\bibitem{song2010modelling}
C.~Song, T.~Koren, P.~Wang, and A.-L. Barab{\'a}si.
\newblock Modelling the scaling properties of human mobility.
\newblock {\em Nature Physics}, 6(10):818--823, 2010.

\bibitem{song2010limits}
C.~Song, Z.~Qu, N.~Blumm, and A.-L. Barab{\'a}si.
\newblock Limits of predictability in human mobility.
\newblock {\em Science}, 327(5968):1018--1021, 2010.

\bibitem{toole2012inferring}
J.~L. Toole, M.~Ulm, M.~C. Gonz{\'a}lez, and D.~Bauer.
\newblock Inferring land use from mobile phone activity.
\newblock In {\em Proc. of ACM SIGKDD}, pages 1--8, 2012.


\bibitem{yongli2015hotplant}
H.~Wang, J.~Ding, Y.~Li, P.~Hui, J.~Yuan, and D.~Jin.
\newblock Characterizing the spatio-temporal inhomogeneity of mobile traffic in large-scale cellular data networks.
\newblock In {\em Proc. of ACM HOTPOST}, pp.~19-24, 2015.

\bibitem{Zhang2012Understanding}
k.~A. Ying~Zhang.
\newblock Understanding the characteristics of cellular data traffic.
\newblock In {\em ACM SIGCOMM CellNet Workshop}, 42(4):13--18, 2012.

\bibitem{yuan2012discovering}
J.~Yuan, Y.~Zheng, and X.~Xie.
\newblock Discovering regions of different functions in a city using human
  mobility and pois.
\newblock In {\em Proc. ACM SIGKDD}, pages 186--194, 2012.

\bibitem{zheng2008understanding}
Y.~Zheng, Q.~Li, Y.~Chen, X.~Xie, and W.-Y. Ma.
\newblock Understanding mobility based on gps data.
\newblock In {\em Proc. of ACM UbiComp}, pages 312--321, 2008.

\end{thebibliography}

\end{document}